\documentclass[hyper,11pt,letterpaper]{JHEP3}

\usepackage{graphicx}
\usepackage{amsmath, amssymb}
\usepackage{amsfonts}
\usepackage{indentfirst}\usepackage[enableskew]{youngtab}
\usepackage{slashed} 
\usepackage{amsbsy} 

\newcommand{\ba}{\begin{align}}
\newcommand{\ea}{\end{align}}
\newcommand{\bea}{\begin{eqnarray}}
\newcommand{\eea}{\end{eqnarray}}
\newcommand{\be}{\begin{eqnarray}}
\newcommand{\ee}{\end{eqnarray}}
\newcommand{\nn}{\nonumber}
\newcommand{\bn}{\begin{enumerate}}
\newcommand{\en}{\end{enumerate}}

\newcommand{\ie}{{\it i.e.~}}
\newcommand{\one}{\mathbf{1}}

\newcommand{\PH}{P_{\CH_\phi}}

\def\half{\frac{1}{2}}

\def\IC{\mathbb{C}}

\def\IN{\mathbb{N}}

\def\IR{\mathbb{R}}
\def\IZ{\mathbb{Z}}

\def\CA{{\cal A}}

\def\CC{{\cal C}}

\def\CE{{\cal E}}
\def\CF{{\cal F}}

\def\CH{{\cal H}}

\def\CL{{\cal L}}
\def\CM{{\cal M}}
\def\CN{{\cal N}}
\def\CO{{\cal O}}

\def\CV{{\cal V}}
\def\CW{{\cal W}}

\def\a{\alpha}
\def\b{\beta}

\def\e{\epsilon}

\def\th{\theta}

\def\m{\mu}

\def\G{\Gamma}

\def\Tr{{\rm Tr}}

\def\Ch{{\rm Ch}}
\def\Ind{{\rm Ind}}
\def\Td{{\rm Td}}

\title{Towards a 4d/2d correspondence for Sicilian quivers}

\author{Lotte Hollands\footnote{hollands@caltech.edu}\ , Christoph
  A.~Keller\footnote{ckeller@theory.caltech.edu}\ , Jaewon Song\footnote{jaewon@theory.caltech.edu}
\\
\\
California Institute of Technology, Pasadena, CA 91125, USA}

\abstract
{
We study the 4d/2d AGT correspondence between four-dimensional instanton
counting and two-dimensional conformal blocks for generalized $SU(2)$
quiver gauge theories coming from punctured Gaiotto curves of
arbitrary genus. We propose a conformal block description
that corresponds to the elementary $SU(2)$ trifundamental half-hypermultiplet, and check it
against $Sp(1)-SO(4)$ instanton counting.   
}

\preprint{
CALT-68-2843
}

\begin{document}

\addtolength{\parskip}{.5mm}
\addtolength{\baselineskip}{.2mm}
\addtolength{\abovedisplayskip}{.8mm}
\addtolength{\belowdisplayskip}{.8mm}

\section{Introduction}

The Alday-Gaiotto-Tachikawa correspondence \cite{AGT} relates 
the instanton partition functions of $\CN=2$
gauge theories to chiral blocks.
Such theories can be engineered in M-theory by wrapping M5-branes
on a punctured Riemann surface $\CC$, the so-called
Gaiotto curve \cite{Gaiotto}. The correspondence is obtained
by decomposing $\CC$ into pairs of pants, and computing
the corresponding conformal block.

If the
corresponding Gaiotto curve $\CC$ is of genus zero or one, a weakly coupled
description exists in certain regions of its complex structure moduli
space. The gauge theory is then described by a linear or cyclic
quiver, and it corresponds to a linear or cyclic 
decomposition of the Gaiotto curve. For these cases
the correspondence between conformal blocks and
instanton partition functions has been checked extensively
in the literature \cite{AGT,Wyllard:2009hg,HKS}.

For Sicilian gauge theories of $A_n$-type with $n \geq 2$, however, 
such a
Lagrangian description does not exist in general.
When decomposing $\CC$ into pairs of pants 
we encounter various strongly interacting isolated
four-dimensional SCFT's whose flavor symmetries are partially
gauged \cite{Gaiotto,Chacaltana:2010ks}, such as the ones
that appear in the Argyres-Seiberg duality \cite{Argyres:2007cn}.

For $A_1$ theories the situation is better:
Sicilian gauge theories of $A_1$-type admit a Lagrangian
description throughout the complex structure moduli space of the
Gaiotto curve $\CC$.  In this case, a pair of pants just refers to a
trifundamental half-hypermultiplet in the four-dimensional gauge
theory.    
Since instanton counting is formally developed for any
$\CN=2$ gauge theory with a Lagrangian prescription, it should be
possible to write down instanton partition functions for Sicilian
$SU(2)$ quiver gauge theories. 
Moreover on the CFT side it should then be possible to write down
conformal blocks corresponding to such decompositions.
Having done this, we should be able to compare the two and check
if the AGT correspondence still works in this case.
To our knowledge this has not yet been accomplished
in the literature, and only indirect checks in
this direction have been performed \cite{Bonelli:2010gk}.

Unfortunately, this extension to Sicilian
quivers is not that straightforward, and introduces
a number of subtleties on both sides of the story.

First of all, note that the conventional method to count instantons in
$SU(2)$ gauge theories is to consider instanton counting for gauge
group $U(2)$, and impose the tracelessness
condition at the end to reduce to $SU(2)$. This
$U(2)$ prescription  
follows from resolving the 
ultra-violet singularities of the instanton moduli space by turning on
an FI parameter. Mathematically, it can be elegantly formulated in
terms of counting rank two torsion-free sheaves.

It is important however to realize that Sicilian quivers are in
general \emph{not} defined for gauge group $U(2)$.
The reason for this is
that the trifundamental 
field, that couples three $SU(2)$ gauge groups, is described by a
half-hypermultiplet. (A full trifundamental hypermultiplet contains
too many degrees of freedom, so that the resulting
theory would not be conformal.\,\footnote{One
  exception to this is the genus two quiver, which describes three $SU(2)$
  gauge groups coupled by two $SU(2)$ trifundamental
  half-hypermultiplets, and can be equivalently 
  described by a full $U(2)$ trifundamental hypermultiplet. We will
  come back to this example in section~\ref{sec:examples}.})
Gauge theories involving half-hypermultiplets are
only CPT invariant if the corresponding matter fields transform in a
pseudo-real representation. (We will present a detailed explanation
of this in section~\ref{sec:gauge}.)  
The fundamental representation of $SU(2)$ is indeed 
pseudo-real, whereas the fundamental of $U(2)$ 
is complex. It is therefore not possible to just work    
with the $U(2)$ as before and specialize to the $SU(2)$
case in the end.

On the face of it there is one obvious way around 
this problem. Instanton
counting for $Sp(N)$ and $SO(N)$ gauge groups has been developed in
\cite{NekrasovShadchin}, and the $D$-type quivers
relevant in this context were investigated in
\cite{Tachikawa:2009rb}.\,\footnote{See also
\cite{Chacaltana:2011ze} for a discussion
of such theories.}
We can thus try to use the fact that
$Sp(1)=SU(2)$ to circumvent those issues
and compute the $SU(2)$ instanton parition function directly.
Similarly, we can also use $SO(4)=SU(2)\times SU(2)$
to compute configurations involving bifundamentals
of $SU(2)$.

This approach was explored in \cite{HKS}. Rather
surprisingly it was found that the
$Sp(1)$ and the $U(2)$ 
instanton counting schemes yield seemingly different results for conformal
quiver gauge theories, even though by the above remarks they
should describe the same physics. The resolution
of this apparent contradiction is that
the instanton partition functions are a priori expressed
in terms of microscopic coupling constants.
To compare the results, they need to be expressed
first in terms of the physical IR coupling, after
which they indeed agree.
To put it another way, the results in the UV are 
related by a non-trivial mapping of
microscopic coupling constants, and thus by a different choice of
renormalization scheme for the conformal gauge
theory. 
For theories that have a string theory embedding,
this mapping has 
an elegant geometric interpretation in terms of 
the corresponding Gaiotto curves.\,\footnote{Geometrically,
  the physical IR coupling corresponds to the period matrix of the
  Seiberg-Witten curve, which is a branched covering over the Gaiotto
  curve. Inequivalent coverings lead to non-trivial mappings of the
  microscopic UV
  coupling constants.  Mathematically, such a UV-UV mapping defines an
  isomorphism between Hitchin systems with the same spectral
  curve.} 

\begin{figure}[htbp]
\begin{center}
\includegraphics[width=1.75in]{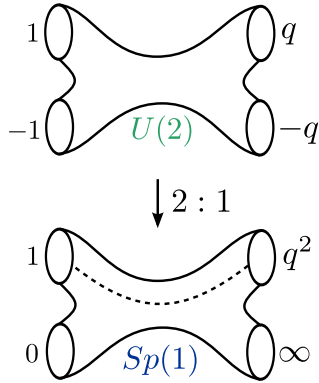}
\caption{The double covering of the the $U(2)$ Gaiotto curve over the
  $Sp(1)$ Gaiotto curve relates the microscopic coupling $q_{U(2)}$
  (which is the cross-ratio of the $U(2)$ Gaiotto curve) to the
  microscopic coupling $q_{Sp(1)} = 4q$. Note that, whereas a $U(2)$
  gauge group is represented by a tube, an $Sp(1)$ gauge
  group is represented geometrically by a tube with a twist-line. In
  the double covering this twist-line gets the interpretation of a
  branch-cut. }\label{Fig:Sp1U2Gcurve}
\end{center}
\end{figure}

Let us briefly recapitulate one of the examples of \cite{HKS}, the
conformal $SU(2)$ gauge theory coupled to four hypermultiplets. The
$Sp(1)$ and $U(2)$ instanton partition functions are related by the
mapping\,\footnote{We have checked this up to order 6 in the microscopic
  couplings.} 
\begin{align}
q_{U(2)} = q_{Sp(1)} \left(1+
        \frac{q_{Sp(1)}}{4}\right)^{-2},
\end{align}
where $q_{U(2)}$ is the microscopic coupling for the $U(2)$ scheme and
$q_{Sp(1)}$ that of the $Sp(1)$ scheme. Geometrically, this mapping
identifies the cross-ratios of the corresponding $U(2)$ and $Sp(1)$
Gaiotto curves, which are related to each other by a double cover
construction. This is illustrated in figure~\ref{Fig:Sp1U2Gcurve}.

Another way of phrasing all this is that the Nekrasov-Shadchin method of
instanton counting singles out a particular choice of coordinates on the
complex structure moduli space of the Gaiotto curve. 
For the AGT correspondence to work, it is clear that
we need to take the same coordinates on the moduli space
of the conformal block as well. 
For standard quiver theories, the $U(2)$ parametrization 
of the moduli space corresponds to the standard choice
of CFT coordinates for the punctured sphere and torus.
The two sides thus agree immediately.
For linear and cyclic $Sp/SO$ quivers, the choice of coordinates
is slightly more involved \cite{HKS}, but has a
very natural interpretation too.
For Sicilian quivers it is no longer obvious
what coordinates to pick, and in fact most
choices have unappealing features, as we 
will discuss below.

Let us emphasize that these complications only arise for conformal
quiver gauge theories, and not for asymptotically free ones. 
For asymptotically free theories, one can \emph{not} have a non-trivial
mapping of parameters because the instanton expansion parameter $q =
\Lambda^{b_0}$ is dimensionful, where $b_0$ is the coefficient of the
beta function. So the instanton counting for all possible realizations of
an asymptotically free gauge theory (such as for a single
trifundamental field coupled to three $SU(2)$ gauge groups)    
should agree directly. On the CFT side this means that the
corresponding correlation function should be essentially
independent of the choice of parametrization.

\begin{figure}[htbp]
\begin{center}
\includegraphics[width=3.95in]{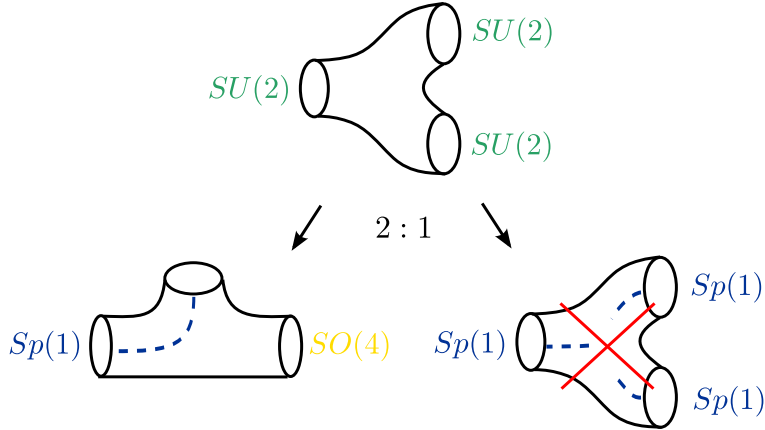}
\caption{The Gaiotto curve for the $SU(2)$ trifundamental as a double
  cover over the $Sp(1)-SO(4)$ Gaiotto curve (on the left) and the
  non-existing $Sp(1)^3$ Gaiotto curve (on the right). On the right is
  illustrated why we cannot find a Gaiotto curve corresponding to the $Sp(1)$
  trifundamental: it is not possible to close the three twist-lines (or
  branch-cuts) on the $Sp(1)$ curve. }\label{Fig:Sp1trifGcurve} 
\end{center}
\end{figure}

After these remarks, let us turn to our main object of interest.  
As we have pointed out, the trifundamental building block 
of Sicilian $SU(2)$ quivers can be
either described as a $Sp(1)$ trifundamental coupling or as a $Sp(1)-SO(4)$
bifundamental coupling. Physically, both of these couplings are
equivalent to an $SU(2)$ trifundamental interaction.
The $Sp(1)-SO(4)$ description has the advantage that it
admits a type IIA string embedding using D4--NS5 branes and
O-branes. Yet 
it only parametrizes a subspace of the moduli space of the $SU(2)$
trifundamental, as it involves just two gauge couplings instead of
three. This is the exact opposite of the $Sp(1)^3$ description, which
cannot be embedded in string theory, but does parametrize the whole gauge
coupling moduli space. This is illustrated in figure~\ref{Fig:Sp1trifGcurve}.   

Whether or not a particular realization of a given $\CN=2$ SCFT has
a string embedding is important when we wish to find its 
dual 2d CFT description. After all, we
expect to find a conformal field theory that lives on the
corresponding Gaiotto curve.

Finding a  conformal field theory prescription for the $SU(2)$
trifundamental half-hyper is conceptually easy.
The general rule is that punctures coming from hypers 
correspond to insertions
of primary fields, and punctures coming from the cutting of
tubes correspond to insertions of descendant fields.
The building block corresponding to a bifundamental
is thus the correlator of one primary and two
descendant fields. 
It is thus
natural to expect that trifundamental building block
should
be a three-point function with three 
descendant fields inserted
\begin{align}
\langle V(\phi^1_{I_1},z_1) V(\phi^2_{I_2},z_2)
V(\phi^3_{I_3},z_3)\rangle\ .
\end{align} 
Since the half-hyper is massless and does not give 
a puncture in the Gaiotto curve, it makes sense
not to insert a corresponding vertex operator.
In our notation the field $\phi^i_{I_i}$ is a Virasoro descendant, and
$V(\phi^i_{I_i},z_I)$ is the vertex operator 
of the field $\phi^i_{I_i}$ inserted at position
$z_i$. The weights of the fields $\phi^i_{I_i}$ encode the Coulomb
branch parameter of an $SU(2)$ gauge group.

The main issue here is the choice of the insertion
points $z_i$. For the standard choice of coordinates
for linear decompositions they turn out as $0,1,\infty$.
For Sicilian decompositions it is no longer clear
what the correct choice is. More precisely, different
choices give different parametrizations of the moduli
space. If we want to match conformal block
to a given instanton counting result, we need to
make sure to pick the correct prescription.

\begin{figure}[htbp]
\begin{center}
\includegraphics[width=4.6in]{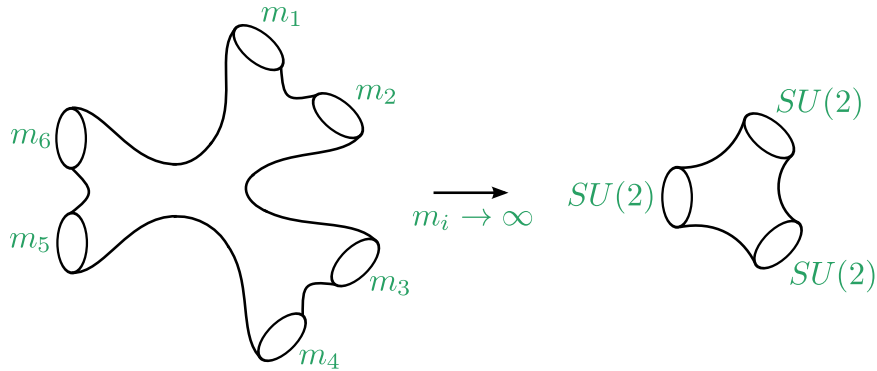}
\caption{On the left is illustrated the Gaiotto curve for the
  conformal quiver
  gauge theory with three $SU(2)$ gauge groups that are all coupled by a
  trifundamental interaction and each individually to two
  massive fundamental hypermultiplets. In the decoupling limit where we take
  all masses $m_i$ of the fundamental hypermultiplets to infinity, we
  are left with the Gaiotto curve corresponding to the asymptotically
  free quiver gauge theory that couples the three $SU(2)$ gauge
  groups by a trifundamental half-hypermultiplet. }\label{fig:cfttrif} 
\end{center}
\end{figure}

Alternatively, we can also be less ambitious and simply check
that the results agree once we express them in terms of
IR variables (\ie as objects defined on the Seiberg-Witten curve). In
particular we can circumvent this entire issue 
by considering asymptotically free theories, where
the relation between UV and IR is trivial.
In this way we can find the conformal block dual to the
$SU(2)$ trifundamental half-hypermultiplet.

We can either start from the conformal quiver
theory with six massive fundamental hypermultiplets
and take a decoupling limit
in which we send all the six masses to infinity. This limit decouples
the hypermultiplets and leaves the three $SU(2)$ gauge groups coupled by a
single trifundamental half-hypermultiplet, see
figure~\ref{fig:cfttrif}. 
Alternatively, 
we can compute the conformal block
as the correlation function of three Gaiotto states,
\begin{equation}\label{eqn:confblocktrif}
\boxed{\langle h_1, \Lambda_1 | V( | h_2, \Lambda_2 \rangle, 1) |  h_3,
\Lambda_3 \rangle} \, 
\end{equation} 
where the Gaiotto state $| h, \Lambda \rangle$ is an eigenstate of
$L_1$ with eigenvalue $\Lambda$. The state  $|h, \Lambda \rangle$ has
appeared in the dual 
conformal field theory description of asymptotically free linear
quiver theories, where it describes the asymptotic boundary conditions
of the quadratic differential on the Gaiotto curve \cite{Gaiotto:2009ma}. 

Finding the conformal block (\ref{eqn:confblocktrif})  dual to the
$SU(2)$ half-trifundamental is one of the main results of this
paper. We verify this prescription by checking several consistency
requirements and by comparing it with instanton counting using the
$Sp(1)-SO(4)$ scheme. We furthermore propose a prescription for the
4d/2d correspondence for any Sicilian quiver, and check this in several
examples.   

The outline of this paper is as follows: We start in
section~\ref{sec:gauge} with an introduction 
to half-hypermultiplets and particularly to instanton counting for
half-hypermultiplets. The result of this exercise is a contour
integral for the $Sp(1)-SO(4)$ bifundamental half-hypermultiplet.
We continue in section~\ref{sec:CFT} 
with the dual conformal field
theory prescription and 
find the three-point function (\ref{eqn:confblocktrif}). In
section~\ref{sec:proposal} we formulate
 the 4d/2d correspondence between conformal blocks and generalized $A_1$
quivers. We check this proposal in several examples in section
\ref{sec:examples}.  
Here we pay special attention to 
conformal quivers, where we find a non-trivial mapping of microscopic
couplings. We conclude and discuss several directions for further research in
section~\ref{sec:discussion}.  
Appendix~\ref{app:flavorenh} contains some more
background on the $SU(2)$ trifundamental half-hyper, whereas
appendix~\ref{app:contourintegrands} summarizes the relevant contour
integrands for Sicilian quivers.

\section{Instanton counting for half-hypermultiplets}\label{sec:gauge} 

The trifundamental fields that appear in Sicilian quiver gauge theories of
type $A_1$ form half-hypermultiplet representations of the $\CN=2$ SUSY
algebra, in contrast to the more common (full) hypermultiplets. In this
section we review the basic properties of Sicilian 
quivers with trifundamental half-hypermultiplets and show that they preserve $\CN=2$ supersymmetry. Subsequently, we develop the tools for counting
instantons in quiver gauge theories with half-hypermultiplets. We
apply these tools in section~\ref{sec:examples} to compute the
instanton partition functions of Sicilian quivers.

\subsection{Half-hypermultiplets}

There are two types of $\CN=2$ supersymmetry multiplets: the $\CN=2$ vector
multiplet  and the hypermultiplet. The former consists of a vector
field $A_\m$, two Weyl fermions $\lambda_\a$ and $\psi_\a$, and one
complex scalar $B$. All of them transform in the adjoint
representation of the gauge group. In $\CN=1$ language such an $\CN=2$
vector 
multiplet consists of one $\CN=1$ vector multiplet with component
fields $(A_\m, \lambda_\a)$ and one chiral multiplet with components
$(B, \psi_\a)$. A 
hypermultiplet requires a choice of representation $R$ of 
the gauge group. In $\CN=1$ language it consists of two chiral
multiplets $Q$ and $\tilde{Q}$, the former transforming in the
representation $R$ and the latter in its complex conjugate $R^*$.  
The chiral multiplet $Q$ has component fields $(\phi,\chi_{\a})$, and
the anti-chiral multiplet $\tilde{Q}$ has components  
$(\tilde{\phi},\tilde{\chi}_{\a})$. 
We call both $Q$ and $\tilde{Q}$ half-hypermultiplets.     

The half-hypermultiplets $Q$ and $\tilde{Q}$ form massless
representations of the $\CN=2$ SUSY algebra. However, even though the
helicities of the states in a half-hypermultiplet form a CPT complete
distribution, the half-hypermultiplet does not transform as a real representation of the
SUSY algebra. Indeed, notice that the massless $\CN=2$ SUSY algebra 
is equal to the Clifford algebra $\mathrm{Cl}_{4, 0}$ with invariance
group $SO(4)$. The four-dimensional representation of the Clifford
algebra $\mathrm{Cl}_{4, 0}$, under which the half-hypermultiplet
transforms, is pseudo-real instead of (strictly) real. A generic
half-hypermultiplet will therefore not be invariant under CPT.  

It is possible though to circumvent this constraint. More precisely,
an $\CN=2$ multiplet is CPT invariant if it transforms under a real
representation of the product of the SUSY algebra, the gauge group and
possible flavor groups. So, apart from the obvious possibility of
combining a chiral and an anti-chiral multiplet into a full
hypermultiplet, we can also consider a single half-hypermultiplet in a
pseudo-real representation of the gauge group $G$.

Nevertheless, there is an additional requirement. Even when a single
half-hyper-multiplet transforms in a real representation of
$\textrm{Cl}_{4,0} \times G$, such a theory may still be anomalous due
to Witten's anomaly argument \cite{Witten:1982fp}. According to
this argument for example a single half-hypermultiplet in the fundamental
representation of $SU(2)$ is anomalous (its partition function
vanishes) since it contains an odd number
of chiral fermions. On the contrary, a single
half-hypermultiplet in the fundamental representation of $SU(2)^3$
contains four chiral fermions in each
$SU(2)$-representation. Therefore, quiver gauge theories with $SU(2)$
trifundamental half-hypermultiplets are free of  Witten's $SU(2)$ anomaly as
well as CPT invariant.

Other examples of consistent theories with half-hypermultiplets occur
when we consider massless bifundamental couplings between $SO$ and
$Sp$ gauge groups. A half-hypermultiplet transforming under the
bifundamental of $SO \times Sp$ is in a pseudo-real representation of
the gauge group $G=SO \times Sp$ and is free of the Witten anomaly as well. 

\subsubsection*{Half-hypermultiplet as a constrained hypermultiplet}

Since working with full hypermultiplets is often much more efficient
than with half-hypermultiplets, in what follows we find an alternative
method to deal with half-hypers. Instead of 
considering a half-hypermultiplet by itself, we start with a full
hypermultiplet (consisting of two half-hypermultiplets) and impose a
constraint on it which only leaves 
a half-hypermultiplet.

A full hypermultiplet can be thought of as a multiplet formed out of
two $\CN=1$ chiral multiplets $Q$ and $\tilde{Q}$. 
The chiral multiplet $Q$ transforms 
in representation $R$ and the anti-chiral multiplet $\tilde{Q}$ in its complex
conjugate $R^*$. 
By the remarks above, for the theory of a single
half-hyper to make sense, $R$ needs to be a pseudoreal.
Let $\sigma_G$ be the anti-linear involution that maps
the representation $R$ to its complex conjugate $R^*$. Since the
representation $R$ is pseudo-real, it obeys $\sigma_G^2=-1$. 
For example, in the case of the fundamental representation
of $SU(2)$, the involution $\sigma_G$ is given by the $\epsilon$-tensor
$i\sigma_2$. For the trifundamental representation of $SU(2)$ it is
given by the product of three $\epsilon$-tensors, one for
each $SU(2)$ gauge group.  

To impose our constraint, we need a map $\tau$ 
that relates $Q$ to $\tilde{Q}$ and vice versa.
Since $\tilde{Q}$ appears in the complex conjugate,
$\tau$ needs to be anti-linear. Moreover, it needs
to preserve the representation, which means
that it must involve $\sigma_G$. 
Let us write a full hypermultiplet as $Q^a$ whose two half-hypermultiplet components are given by $Q^1 = Q$ and $Q^2=\tilde{Q}^*$. 
The involution $\tau$ is then defined by 
\be
Q^a \mapsto \tau(Q^a) = \sigma_G\otimes \sigma_I (Q^a)^*\,,
\ee
where
\be
\sigma_I \left(\begin{array}{c}Q^1\\ Q^2\end{array}\right)
= \left(\begin{array}{r}-Q^2\\ Q^1\end{array}\right)\,.
\ee 
It is straightforward to check that indeed $\tau^2=1$.
%The involution $\tau$ is real as $\tau^2=1$. 
We can describe a half-hypermultiplet as a 
hypermultiplet that stays (anti-)invariant under $\tau$,
\ie that is an eigenvector of $\tau$ of eigenvalue $\pm1$.
More explicitly, such a hyper is given by $(Q,\pm \sigma_G Q^*)$.

This description of a half-hypermultiplet is for instance
convenient
to find the Lagrangian for a half-hyper $Q$ starting from the Lagrangian
of a full hyper. 
Recall that the Lagrangian for the hypermultiplet $Q^a=(Q,\tilde{Q}^*)$ 
coupled to a vector multiplet $(V,\Phi)$ is given by
\begin{align}
\CL_{\textrm{fh}} &=  \int d^2 \th \, d^2 \bar{\th} \left( Q^{\dag} e^{V}
Q + \tilde{Q}^t e^{-V} \tilde{Q}^{*} \right) + 
 2 \sqrt{2}~\textrm{Re} \int d^2 \th \left( \tilde{Q}^t \, \Phi \,
  Q  \right).    \notag                        
\end{align}
For pseudo-real representations
we can apply the constraint $\tilde{Q} = \pm \sigma_G Q$ to recover the Lagrangian 
\begin{align}
\CL_{\textrm{hh}} &= \int d^2 \th \, d^2 \bar{\th} \left( Q^{\dag}
  e^{V} Q \right) \pm  
\sqrt{2}~\textrm{Re} \int d^2 \th \left( Q^t \sigma_G^t \, \Phi \,
  Q  \right)    \notag                        
\end{align}
for a single half-hypermultiplet. Here we rescaled $Q \to \frac{1}{\sqrt{2}} Q$ to
give the kinetic term in the Lagrangian a canonical coefficient. We
also used 
\begin{align*}
\tilde{Q}^t e^{-V} \tilde{Q}^{*} = Q^t \sigma^t_G e^{-V} \sigma_G Q^*
= Q^t e^{V^t} Q^* = Q^{\dag} e^{V} Q \,,
\end{align*}
since $\sigma_G^{-1} T \sigma_G = -T^t$ for 
$T \in \mathfrak{g}$. 
Since we found the Lagrangian $\CL_{\mathrm{hh}}$ by starting out with
the Lagrangian $\CL_{\mathrm{fh}}$ for a full hypermultiplet and then
applying the constraint  $\tilde{Q} = \pm \sigma_G Q$, it is
automatically invariant under $\CN=2$ supersymmetry. 

Let us spell this out in some more detail. Substituting the constraint $\tilde{Q} =
\pm \sigma_G Q$ in the $\CN=2$ supersymmetry equations yields two
identical copies of the supersymmetry variations for the components of
$Q$, which depend on all of the $\CN=2$ supersymmetry parameters. The
Lagrangian $\CL_{\rm hh}$ is obviously invariant under these
variations. The
$SU(2)_R$ symmetry now acts on the vector of complex scalars $(q,\pm
\sigma_G q^*)^t$.

As an example, the Lagrangian for the trifundamental $SU(2)$
half-hypermultiplet reads in components 
\begin{align}
\CL_{\textrm{trif}} &= \int d^2 \th \, d^2 \bar{\th} \left(
  Q^*_{abc} e^{(V_1)^{a}_{~a'}} Q^{a'bc} +   Q^*_{abc}
  e^{(V_2)^{b}_{~b'}} Q^{ab'c} +   Q^*_{abc} e^{(V_3)^{c}_{~c'}} Q^{abc'} \right) \\ &~\pm  \sqrt{2} ~\textrm{Re} \int d^2 \th \left( \e^{bb'} \e^{cc'}  Q_{abc} \, \Phi^{aa'} \,
  Q_{a'b'c'} +  \e^{aa'} \e^{cc'}  Q_{abc} \, \Phi^{bb'} \,
  Q_{a'b'c'}  +  \e^{aa'} \e^{bb'}  Q_{abc} \, \Phi^{cc'} \,
  Q_{a'b'c'}   \right).    \notag                        
\end{align}
We can obtain the bifundamental hyper by demoting one of gauge groups
to a flavor group. From this perspective it is clear that the
bifundamental has an enhanced 
$SU(2)$ flavor symmetry, as we already knew from general
principles. We discuss these and other aspects of the $SU(2)$
trifundamental in detail in appendix~\ref{app:flavorenh}.

\subsection{Instanton counting for half-hypermultiplets}

We now turn to instanton counting for
half-hypermultiplets. Also for this purpose it is convenient to use the
description of a half-hypermultiplet as a constrained full
hypermultiplet. Instanton counting for any $\CN=2$ gauge theory with
full hypermultiplets is developed in \cite{Nekrasov:2002qd, NekrasovShadchin}
and spelled out in more detail in for example \cite{Shadchin:2005mx,HKS}. It is performed by
topologically twisting the $\CN=2$ gauge theory. The resulting instanton
partition function is given by the integral 
\begin{align}\label{eqn:Zinst}
    Z^{\rm inst} = \sum_k q_{\rm UV}^k \oint_{\CM^{G}_k} e(\CV)
\end{align}
over the ADHM moduli space of instantons $\CM^{G}_k$ for the gauge
group $G$ and instanton number $k$, where the Euler class $e(\CV)$
encodes the matter content of the gauge theory. More precisely, the
vector bundle $\CV$ is equal to the space of solutions to the Dirac
equation for the chosen matter representation in the self-dual instanton
background.    

Let us emphasize that the $SU(2)$ R-symmetry is essential for
performing the topological twist. We identify the new Lorentz group of
the twisted $\CN=2$ theory as
\begin{align*}
L' = SU(2)_L \times
\mathrm{diag}(SU(2)_R \times SU(2)_I)\ ,
\end{align*}
where we denoted 
the R-symmetry group by $SU(2)_I$ to avoid confusion. 
After twisting
the two complex scalars of a full hypermultiplet combine into a Weyl
spinor 
\begin{align*}
\Psi= (\psi_1, \psi_2) = (q, \tilde{q}^*)\ ,
\end{align*}
\ie the R-symmetry index
turns into a spinor index. The matter part of the theory
localizes to solutions of the Dirac
equation\,\footnote{Although we write down an explicit form of the
  Dirac equation for a spinor transforming in the fundamental
  representation of a single gauge group, equation~(\ref{Weyleq}), as
  well as the following equations,
  should be read abstractly and can easily be adapted to hold in
  a generic setting.} 
\be \label{Weyleq}
(i\sigma^\mu\partial_\mu + \sigma^\mu A_\mu)\Psi =0
\ee 
in the self-dual instanton background determined by the gauge field
$A$, with $\sigma^\mu=(\mathbf{1},i\sigma^i)$ 
(note that we are in Euclidean signature). These solutions form a
vector bundle over the moduli space of self-dual instantons,
localizing the path-integral to the integral over the moduli space of instantons (\ref{eqn:Zinst}). 
Actually computing the instanton partition function~(\ref{eqn:Zinst}) can then be reduced
to evaluating the 
equivariant index of the Dirac operator with respect to a torus action
on the ADHM moduli space.   

For half-hypermultiplets the twisting works similar, since we have
established the R-symmetry invariance of the half-hypermultiplet
Lagrangian. Let us start with a twisted full hypermultiplet.
Since the R-symmetry indices of the scalars in the full hypermultiplet
turn into spinor indices, 
we can again define the map
\be
\Psi \mapsto \tau(\Psi) = \sigma_G\otimes \sigma_I \Psi^*\ .
\ee
As before, the matrices $\sigma_G$ and $\sigma_I$ act on the gauge and
spinor indices, respectively. In particular,
\be
\sigma_I \left(\begin{array}{c}\psi_1\\ \psi_2\end{array}\right)
= \left(\begin{array}{r}-\psi_2\\ \psi_1\end{array}\right) \notag
\ee 
The path integral of the half-hypermultiplet theory
localizes onto solutions of the Dirac equation (\ref{Weyleq}) that are
invariant under $\tau$. 

The involution $\tau$ indeed maps solutions of
the Dirac equation 
(\ref{Weyleq}) to solutions, as can be seen from 
\be
(i\sigma^\mu\partial_\mu + \sigma^\mu A_\mu)\tau(\Psi)
=-\sigma_G\otimes \sigma_I
((i\sigma^\mu\partial_\mu + \sigma^\mu A_\mu)\Psi)^* = 0\ ,
\ee
where we have used $\sigma_I^{-1}\sigma^\mu\sigma_I = (\sigma^{\mu})^*$
and $\sigma_G^{-1}A_\mu \sigma_G = - A_\mu^*$. 
We can thus find a basis of the space of solutions to the Dirac
equation on which $\tau$ acts with eigenvalue $\pm 1$. The relevant
solutions for the single half-hypermultiplet  are given by those
basis elements 
which all have eigenvalue $+1$ (or all eigenvalue $-1$) under $\tau$,
and form a  
half-dimensional vector bundle over the moduli space $\CM_G^k$ of self-dual
instantons. 

\begin{figure}[htbp]
\begin{center}
\includegraphics[width=2.5in]{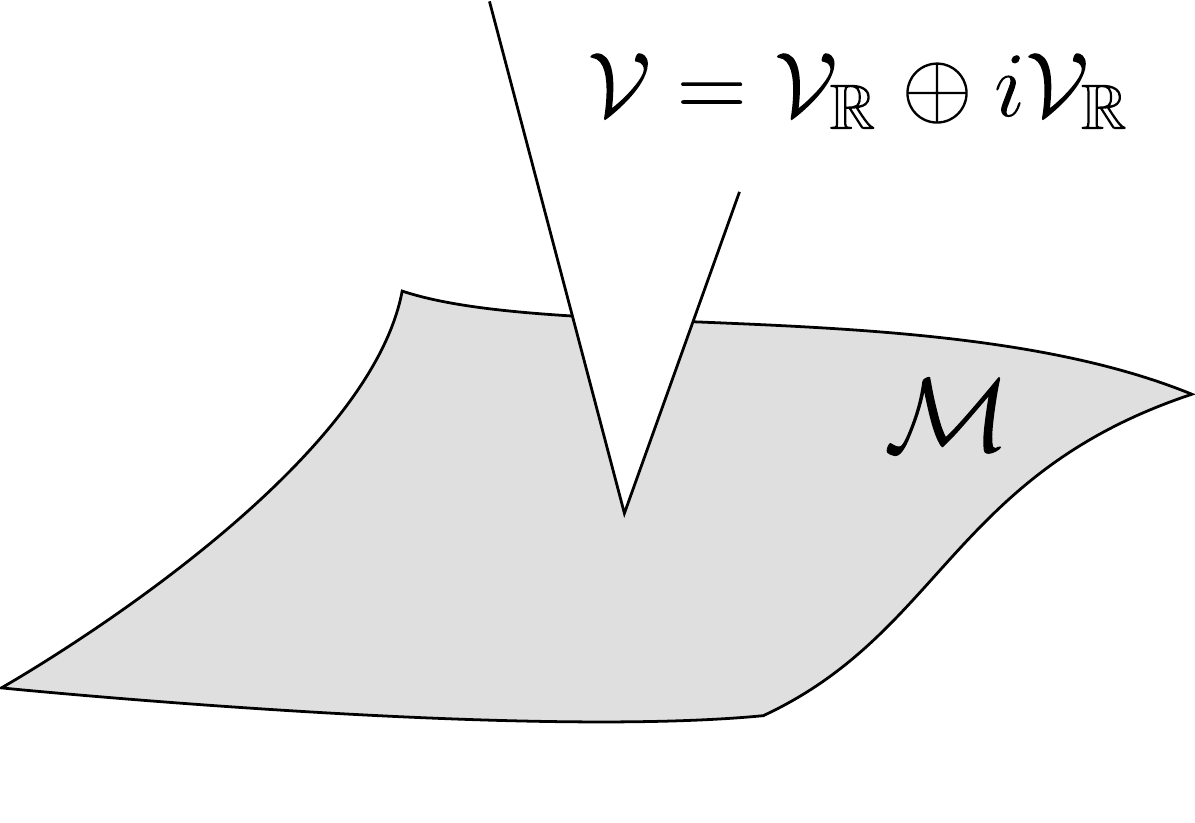}
\caption{The solutions to the Dirac equation in a given
  representation of the gauge group form a vector bundle $\CV$
  over the ADHM moduli space $\CM$. A pseudo-real representation
  induces a real structure $\tau$ on the vector bundle
  $\CV$ that splits it into two copies $\CV = \CV_{\IR} \oplus i
  \CV_{\IR}$. The relevant solutions for a half-hypermultiplet are either parametrized by $\CV_{\IR}$ or 
  $i \CV_{\IR}$.  }\label{Fig:ADHMgeom} 
\end{center}
\end{figure}

As an intermezzo, remember that the space of fermionic solutions to the Dirac
equation in a pseudo-real representation always admits a real
structure. It is not hard to 
see that the anti-linear involution $\tau$ in fact defines this real
structure. So let us consider a basis of
solutions on which $\tau$ acts with eigenvalues $\pm 1$. Whereas for a
theory with a hyper all solutions with eigenvalue $+ 1$ or $-1$ need to be taken into
account, the theory with a half-hyper enforces a restriction
 to the solutions with eigenvalues either all $+1$ or all $-1$. 

Let us name $V$ the total vector space of solutions to the Dirac
equation in a given instanton background. Then the real structure 
$\tau$ induces a splitting 
\be
V = V_R \oplus i V_R\ .
\ee 
The vector space
$V_R$ (called the real form of $\tau$) consists of solutions with
eigenvalue $+1$, whereas $iV_R$ consists of solutions with
eigenvalue $-1$. Indeed, since $\tau$ is anti-linear, multiplying a 
solution $\Psi \in V_R$ by $i$ yields a solution with eigenvalue $-1$.  The real structure $\tau$ reduces the
group of basis transformations acting on $V$ from $U(d)$ to $SO(d)$,
where $d$ is the dimension of $V$. The action of $SO(d)$ leaves
$V_R$ invariant. 

The two half-hypermultiplets that make up a hypermultiplet are defined by
the two constraints $\tilde{Q} = \pm \sigma_G Q$. One
half-hypermultiplet singles out the subspace  $V_R \subset V$, and the
other the subspace $i V_R \subset V$. So multiplying the solutions of
the Dirac equation by $i$ brings us from one half-hypermultiplet to the other. 

\subsubsection*{Instanton partition functions for half-hypermultiplets}

Let us summarize the above. Consider an $\CN=2$ gauge theory
coupled to a full hypermultiplet in a pseudo-real representation of the
gauge group. Its instanton partition function is given by
equation~(\ref{eqn:Zinst}). This is in fact equal to
\begin{align}
    Z_{\rm fh}^{\rm inst} = \sum_k q_{\rm UV}^k \oint_{\CM^{G}_k} e(\CV_R
    \oplus i \CV_R),
\end{align}
since the pseudo-real representation defines a real structure on the
complex vector bundle $\CV$ of solutions to the Dirac equation. The
bundle $\CV_R$ is an oriented real bundle, whose Euler class is
defined as the Pfaffian (this is only non-trivial when the rank of the
bundle is even). The Euler class of its 
complexification $\CV=\CV_R   \oplus i \CV_R$ can then be expressed as
the square of the Euler class of $\CV_R$, 
\be
e(\CV_R   \oplus i \CV_R) = e(\CV_R)^2.
\ee
This equality continues to hold for the equivariant Euler classes
$e_{\mathbf{T}}(\CV)$ and $e_{\mathbf{T}}(\CV_R)$, with respect to the
torus action $\mathbf{T}= T_{a_k} \times T_{\phi_i} \times U(1)_{\e_1,\e_2}$ on the ADHM moduli space, where $T_{a_k}$ is the torus of the gauge group,
$T_{\phi_i}$ the torus of the dual group, and the action of
$U(1)_{\e_1,\e_2}$ on $\IR^4$ defines the Omega-background. 

In other words, when the rank of $\CV_R$ is even (\ie when the complex Dirac
index of the pseudo-real representation is even), the instanton
partition function for a half-hyper theory localizes as  
\begin{align} 
    Z^{\rm inst}_{\rm hh} = \sum_k q_{\rm UV}^k \oint_{\CM^{G}_k} e(\CV_R).
\end{align}
Since the involution $\tau$ commutes with the torus $\mathbf{T}$, we
can compute the contribution of a half-hypermultiplet equivariantly by
just taking the square-root of the product of weights for the full
hypermultiplet theory.

In this manner we can compute the instanton partition function for
the $Sp(1)$ trifundamental half-hypermultiplet and the $Sp(1)-SO(4)$
bifundamental half-hypermultiplet.\,\footnote{This derivation justifies the
  method used in \cite{HKS} to compute the contribution of the
  $Sp(1)-SO(4)$ bifundamental half-hypermultiplet.} Notice that in
both examples the Dirac index is even for any instanton number $k$. In
section~\ref{sec:examples} we apply this scheme to evaluate instanton
partition functions corresponding to $Sp(1)/SO(4)$ quiver gauge
theories. 
%In section~\ref{sec:discussion} we discuss instanton
%counting for the $Sp(1)^3$ trifundamental half-hypermultiplet. 

\section{CFT building blocks for Sicilian quivers}\label{sec:CFT}

Let us now discuss the building blocks that are
needed for the AGT correspondence.
In the correspondence for conformal
$SU(2)$ quiver gauge theories hypermultiplets 
are given by punctures on the Gaiotto curve.
Gluing the neighborhoods of two punctures to create a tube gauges 
the flavor symmetry group of the two hypermultiplets into an 
$SU(2)$ gauge group. The masses of the two hypermultiplets have to be
opposite to perform the gluing, since they correspond
to the residue of the Seiberg-Witten 1-form at the puncture. The masses
then turn into the Coulomb parameters $\pm a$ of the $SU(2)$ gauge group
after the gluing.

On the CFT side, hypermultiplets correspond
to insertions of primary fields $\phi^i$ 
whose conformal weights are related
to the masses of the hypermultiplets. 
A gauge group corresponds
to inserting a complete set of descendants
of a given primary field.
We recall that an arbitrary Virasoro descendant $\phi_I$ at level $N$
is given by a partition $I$ of $N$ by $\phi_I =\prod_j L_{-I_j}\phi$.
For ease of notation we will also just write $I$ for $N$.
The projector on a particular representation
that we insert can thus be written as
\be
\PH = \sum_{I,J} K^{-1}_{IJ} |\phi_J\rangle\langle \phi_I|\ ,
\ee
where $K^{-1}$ is the inverse of
the Kac matrix $(K)_{IJ}= \langle \phi_I|\phi_J\rangle$.
The modulus of the tube corresponds to
the coupling of the gauge group.
From this it is clear that if we decouple the 
gauge group by sending $q\rightarrow 0$ we
recover the original expression for the
ungauged theory, since the contributions
of the descendants vanish and only the primary
field survives.

The complete instanton partition function can thus
be obtained from a pair of pants decomposition
of the Gaiotto curve. Its building blocks
are given by three-point functions containing
one or more descendant fields, and the total
expression is obtained by summing over
all descendant fields in the channels.
This sum corresponds to the sum over the fixed points
in the instanton counting.
For linear and cyclic quivers, 
the only building blocks needed are hypermultiplets
in the fundamental and hypermultiplets in the
bifundamental. The corresponding CFT expressions
are three-point functions with one or two
descendant fields.

\begin{figure}[htbp]
\begin{center}
\includegraphics[width=5.6in]{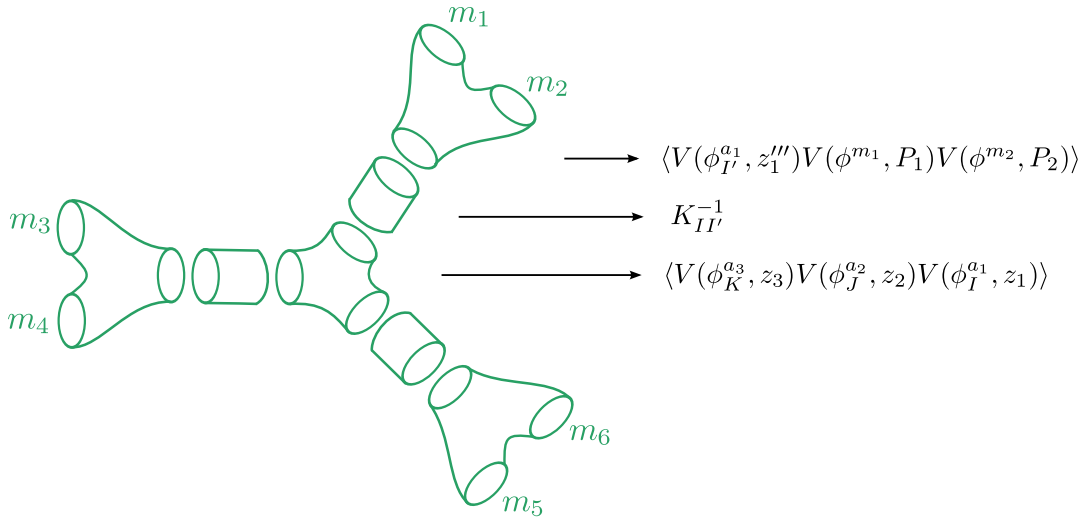}
\caption{Decomposition of the sphere with six punctures
  into three-punctured spheres and tubes, and the corresponding
  conformal blocks.}\label{Fig:CFTtrifdecomposed}
\end{center}
\end{figure}

For Sicilian quivers such as in 
figure~\ref{Fig:CFTtrifdecomposed},
however, we also need hypers in the
fundamental of three different gauge groups.
The corresponding CFT building block should then
be described by the three-point function with
three descendant fields inserted, 
\be \label{threedescendants}
\langle V(\phi^1_{I_1},z_1) V(\phi^2_{I_2},z_2)
V(\phi^3_{I_3},z_3)\rangle\ .
\ee
Here we have used the notation $V(\phi,z)$ for the
the vertex operator corresponding to the field
$\phi$ inserted at $z$.
The weights of the fields $\phi^i$ are
related to the Coulomb branch parameters $a_{1,2,3}$ of the
three $SU(2)$ gauge groups involved.   
Choosing the insertion
points $z_i$ is quite subtle and affects the outcome,
as we will now discuss.

\subsection{Three-point functions}

Let us start with a
reminder about three-point functions and some of their properties.
For three primary fields the three-point function
is fixed up to a constant $C_{123}$. The coordinate
dependence itself is fixed covariance under 
by M\"obius transformations,
\ie the global conformal symmetry.

Local conformal symmetry allows us to compute three-point functions of
arbitrary descendants of those primary fields as well.
In principle, this is straightforward: the only thing
needed is the OPE of the stress energy tensor $T(z)$ with 
the primary fields with itself. We can then use
\be
\langle V(L_{-n}\phi,z)\ldots\rangle =
\oint_{z}dw \, (w-z)^{-n+1}\langle T(w) \, V(\phi,z)\cdots \rangle
\ee
to reduce the three-point function to contour integrals
of the correlator of three primary fields and several
energy stress tensors. This correlator is a meromorphic function on a
Riemann surface and thus determined by its poles.
We can thus consecutively eliminate the $T(z)$ by
summing their OPEs with the other $T(w)$ and 
the primary fields, until we are left with
just the three-point function of the primary fields. 
We can then evaluate the contour integral. 

Though conceptually simple, in practice this
procedure is quite cumbersome. Since most of the
time we are interested in very specific values of
$z_i$ only, it can be more efficient to phrase
the computation in terms of operators on
the Hilbert space of a Virasoro representation.
The operator-state correspondence tells us
that 
\be
\lim_{z\rightarrow0}\phi(z)|0\rangle = |\phi\rangle\,. 
\ee
The corresponding bra state is given by the
operator at infinity. More precisely, it is
obtained from the ket state using the 
M\"obius transformation $z\mapsto 1/z$:
\be
\lim_{z\rightarrow0}\langle 0| V(z^{-2L_0}  e^{-\frac{1}{z}L_1} 
\phi,1/z)=\langle\phi|\,. 
\ee
The three-point function with primary fields
at 0,1,$\infty$ can then be computed as\,\footnote{Strictly speaking we
can only do this for $h_3-h_1\in \IZ$. From general arguments
we know however that the coefficients of the conformal block 
are given by rational functions in $h_i$ and $c$.
The expressions we obtain thus continue to be valid for arbitrary
values of $h$.}
\be
\langle \phi^1|V(\phi^2,1)|\phi^3\rangle = \langle\phi^1|\phi^2_{h_3-h_1}|\phi^3\rangle
= C_{123}\,.
\ee
We can compute such three-point functions with descendant bra
and ket states by commuting through all Virasoro operators
using
\be
[L_n, \phi_m] = (n(h-1)-m)\phi_{m+n} 
\ee 
for primary fields $\phi^2$.
If $\phi^2$ is a descendant, then
we first need to express it in terms of Virasoro operators
and modes of the primary field, which we do by using the following expression for 
the $-N_1$ mode of a $V(L_{-N_2}\phi,z)$ \cite{Gaberdiel:2000qn}
\begin{multline}
V_{-N_1}(L_{-N_2}\phi) 
= \sum_{l\geq 0} \binom{N_2-2-l}{l} L_{-N_2-l} V_{-N_1+N_2+l}(\phi)\\
+(-1)^{N_2} \sum_{l\geq 0} \binom{N_2-2-l}{l} V_{-N_1-l+1}(\phi)
L_{l-1}\,.
\end{multline}
Note that even though the sums are infinite, 
they reduce to finite sums when acting on any
particular state.

Consider a theory with
several identical gauge groups. One would expect that
the partition function
should be symmetric under suitable permutations of
the gauge group. On the CFT side this means that
the three-point function should be symmetric under
permutations of the insertion points.

For instance, if a theory contains a hypermultiplet in the
fundamental of two $SU(2)$ groups, then
the three-point function must be symmetric
under exchanging the two. This is indeed the case, 
as follows from
\be \label{twoSym}
\langle \phi^1_{I_1}|V(\phi^2,1)|\phi^3_{I_3}\rangle
=\langle \phi^3_{I_3}|V(\phi^2,1)|\phi^1_{I_1}\rangle\ .
\ee
To see that (\ref{twoSym}) indeed holds we can use
the M\"obius transformation $z\mapsto 1/z$. A general
field transforms under a M\"obius transformation
$\gamma$ as 
\be \label{Mobius}
V(\phi,z) \mapsto V\left((\gamma(z)')^{L_0}e^{\frac{\gamma''}{2\gamma'}L_1}\phi,\gamma(z)\right)\ .
\ee
From this we see that as long
as $\phi^2$ is a primary field, it does not pick up any
correction terms from this transformation.

On the other hand, if we consider the case 
of a hyper in the fundamental of three gauge groups,
we need to insert three descendants, and $V(\phi^2,1)$ will
no longer transform in such a simple way. The usual
vertex is then no longer symmetric under permutations,
as the M\"obius transformations that exchange punctures
introduce corrections. This means that the standard
CFT vertex must correspond to a regularization scheme
of the gauge theory which treats the gauge groups
differently.

More generally, if we use any M\"obius transformation
to change the insertion points of a three-point function
with descendants, then due to (\ref{Mobius}) we
will pick up corrections. This means that the
detailed expression for the three-point function greatly depends
on the choice of insertion points $z_i$ in (\ref{threedescendants}).
It turns out that these issues are less severe for
asymptotically free theories. Let us therefore
turn to those cases.

\subsection{Partition function for the trifundamental coupling}

Conformal blocks a priori correspond to conformal 
gauge theories, as the flavor symmetries always
work out in such a way that there are four fundamental
hypers per $SU(2)$ gauge group. We can however obtain
asymptotically free theories by sending the mass
of hypers to infinity and so decoupling them.
More precisely, to decouple a hyper of mass $m$
in the fundamental of a gauge group of coupling
$q$ we take
\be \label{masslimit}
q \rightarrow \Lambda/m\ , \qquad m \rightarrow \infty\ .
\ee
Here $\Lambda$ is the scale of the newly asymptotically free
theory.
In this way we can obtain any asymptotically free
partition function from a conformal block.

Let us use the procedure outlined above to compute
the partition function of a half-hypermultiplet
in the trifundamental of $SU(2)$. We start out
with conformal theory which corresponds to a sphere
with six punctures (see figure~\ref{Fig:CFTtrifdecomposed}), but decompose it in a symmetric
(\ie non-linear) way:
\begin{multline} \label{CFT6pct}
Z = \sum_{I_1,I_2,I_3}\sum_{J_1,J_2,J_3} 
\langle \phi^{m_1}|V(\phi^{m_2},1)|\phi^{a_1}_{I_1}\rangle
\langle \phi^{m_3}|V(\phi^{m_4},1)|\phi^{a_2}_{I_2}\rangle
\langle \phi^{m_5}|V(\phi^{m_6},1)|\phi^{a_3}_{I_3}\rangle\\
\times
K^{-1}_{I_1J_1}K^{-1}_{I_2J_2}K^{-1}_{I_3J_3}\,
\langle \phi^{a_1}_{J_1}|V(\phi^{a_2}_{J_2},1)
|\phi^{a_3}_{J_3}\rangle\, q_1^{I_1}q_2^{I_2}q_3^{I_3}\,.
\end{multline}
Note that we have chosen more or less by fiat that
the trifundamental vertex, \ie the three-punctured
sphere in the center of the decomposition, is given
by the sphere with punctures at $0,1,\infty$. In view of
the remarks in the previous section the result is 
certainly not symmetric under permutation of the gauge
groups. 
To obtain the asymptotically free theory, that is
the result for a single half-hyper in the trifundamental,
we apply (\ref{masslimit}). 
It turns out that the resulting expression \emph{is} symmetric
under permutations up to spurious terms (which we explain in a moment). 
It is moreover independent
on the choice of punctures of the three-punctured sphere in the center
of the decomposition, up to a simple rescaling of the couplings $q$.

This rather surprising result can be better understood
when computing asymptotically free theories
using Gaiotto states \cite{Gaiotto:2009ma}.
Such a state $|h,\Lambda\rangle$ is an eigenstate of the Virasoro mode
$L_1$ with  
eigenvalue $\Lambda$,
\be\label{Gaiottostate}
L_1|h,\Lambda\rangle = \Lambda|h,\Lambda\rangle \qquad L_n|h,\Lambda\rangle = 0 \quad n\geq 2\,.
\ee
More concretely such as state can be written as a power series
in $\Lambda$
\be
|h,\Lambda\rangle = \sum_{n=0}^\infty \Lambda^n |v_n\rangle\,,
\ee
where $|v_0\rangle = |h\rangle$ and $|v_n\rangle$ is a specific
linear combination of Virasoro descendants of
$|h\rangle$ at level $n$.
These states can then be used to compute instanton partition
functions for asymptotically free $SU(2)$ theories. The norm
of such a state for instance gives the instanton partition
function of pure $SU(2)$ gauge theory. Both states in this norm
originate from decoupling a pair of hypermultiplets in the conformal
$SU(2)$ gauge theory. The conditions
(\ref{Gaiottostate}) come from the poles of the quadratic
differential $\phi_2(z)$ on the Gaiotto curve.
(See \cite{Marshakov:2009gn} for a proof that this is
equivalent to the infinite mass limit.)

It is natural to use the same strategy also for
multiple gauge groups. The $SU(2)$ trifundamental can be
obtained by decoupling three pairs of hypers in the conformal $SU(2)$
gauge theory corresponding to the six-punctured sphere. We thus
compute the three-point function
\be\label{CFTtrifund}
Z_{\textrm{CFT}}=\langle h_1,\Lambda_1|V(|h_2,\Lambda_2\rangle,1)|h_3,\Lambda_3\rangle\,.
\ee
This gives indeed the same expression
as the one we obtained above.
Now we can also explain why (\ref{CFTtrifund}) is invariant
under permutation of the three gauge groups (up to some
trivial factors).
As usual we use a M\"obius transformation $\gamma$ to
exchange the three insertion points. 
From (\ref{Mobius}) and (\ref{Gaiottostate}) it follows that
the Gaiotto state $|h,\Lambda\rangle$ 
transforms to 
\be
e^{\frac{\gamma''}{2\gamma'}\Lambda} (\gamma')^h |h,\gamma'\Lambda\rangle\ ,
\ee
so that after a redefinition of $\Lambda$ the two 
three-point functions only
differ by a spurious prefactor. Since this holds for
any M\"obius transformation, the result is essentially
independent of the insertion points.

We propose that (\ref{CFTtrifund}) 
is equal to the instanton partition function of a half-hyper in
the trifundamental representation of $SU(2)$ (up to a spurious
factor\,\footnote{In the following we define a spurious factor as a
  factor that does not depend on the Coulomb branch parameters and
  only contributes to the first terms of the genus expansion of the free
  energy.}). Even though we did not 
compute this 
partition function directly, we can perform several
consistency checks on (\ref{CFTtrifund}). First note
that it has indeed a proper $\CF_g$ expansion, \ie
that it can be written
\be
Z = \exp\left(\sum_{g\geq 0} \hbar^{2g-2} \CF_g\right),
\ee
with no higher negative powers of $\hbar$ appearing. Second,
(\ref{CFTtrifund}) reduces correctly to the $SU(2)$ 
bifundamental when we decouple one of the gauge
groups. Finally, when setting $\Lambda_2=\Lambda_3$
it agrees with the partition function 
of a $Sp(1)-SO(4)$ gauge theory with a single
hyper in the bifundamental (details of this check
can be found in section~\ref{sec:examples}).

\section{Towards a 4d/2d correspondence for Sicilian
  quivers}\label{sec:proposal}

The simplest way to define a conformal $\CN=2$ Sicilian 
$SU(2)$ quiver gauge theory is through its M-theory construction.
Wrap two M5 branes on a Riemann surface with punctures $\CC$.
The quiver theory corresponding to a particular duality
frame is obtained from a decomposition of $\CC$ into
pairs of pants. The punctures of $\CC$ correspond
to hypermultiplets, and the tubes connecting the different
pants correspond to $SU(2)$ gauge groups whose
microscopic coupling constants are given by the 
complex structure moduli of the tubes.

\begin{figure}[htbp]
\begin{center}
  \begin{minipage}{.43\textwidth}
      \includegraphics[width=\textwidth]{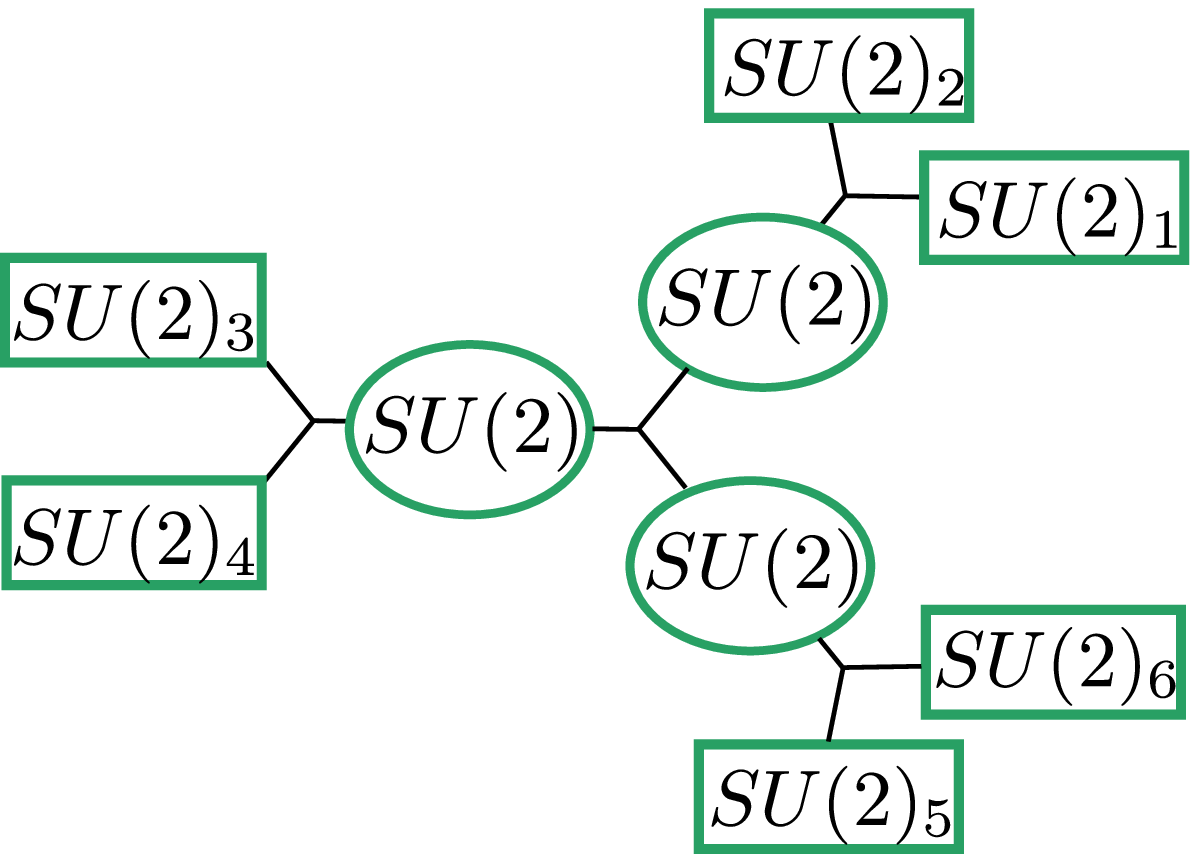} 
   \end{minipage}
 \begin{minipage}{.08\textwidth}
     $\longleftrightarrow$
   \end{minipage}  
 \begin{minipage}{.47\textwidth}
      \includegraphics[width=\textwidth]{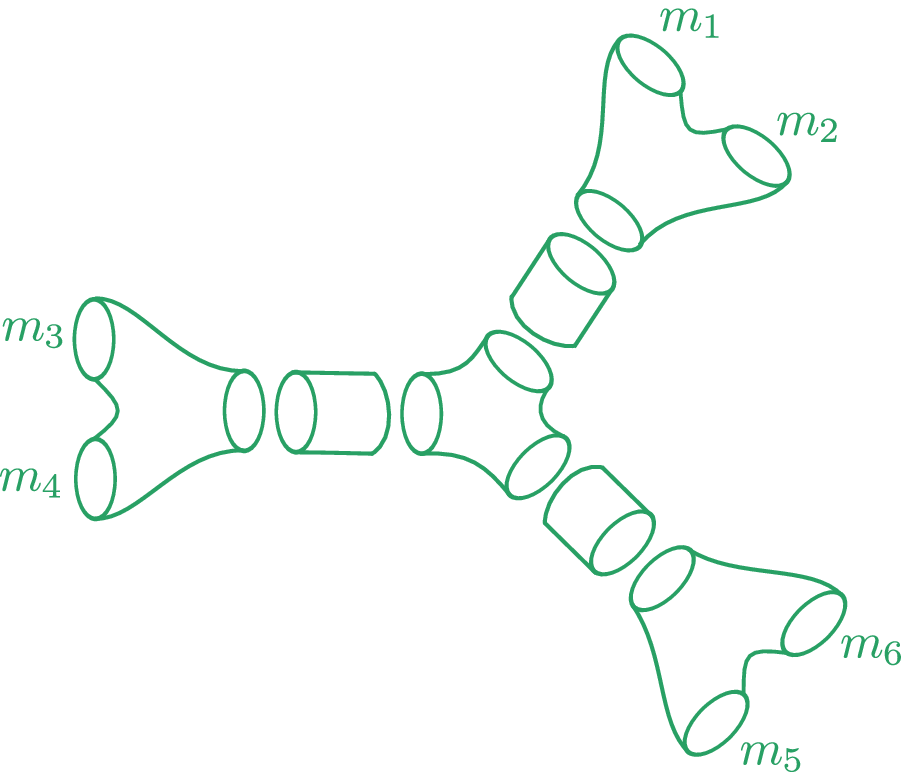}
   \end{minipage} 
\caption{Illustration of the correspondence between instanton partition
  functions of Sicilian $SU(2)$ quiver
  gauge theories and Virasoro conformal blocks on the corresponding
  Gaiotto curve for the six-punctured sphere. Each $SU(2)$ gauge group in
  the quiver is mapped to a tube in the Gaiotto curve, whereas
  $SU(2)$ matter is represented by three-punctured
  spheres.}\label{Fig:SicilianAGT} 
\end{center}
\end{figure}

The building blocks are thus spheres with three punctures
or tubes.
There are three different configurations. The sphere
with two punctures and one tube corresponds to two
hypermultiplets in the fundamental. The sphere
with one puncture and two tubes corresponds to
a hyper in the bifundamental of the two $SU(2)$.
Finally, as a new element, there is the
the sphere with three tubes. It corresponds to
a half-hyper in the trifundamental. Since
the half-hyper is massless, it is natural not
to have a puncture for it in this building block. See
figure~\ref{Fig:SicilianAGT} for an example.

Quivers with asymptotically free gauge groups can
always be obtained from conformal theories
by sending the mass of one of the hypers
to infinity.

We can then compute the conformal block 
for this quiver in the following way.
First, at every puncture insert a primary
field whose conformal weight is given
by the mass of the hyper in the usual way.
Second, for every tube insert a projector
\begin{align*} 
\PH = \sum_{I,J} q^{|I|} K^{-1}_{IJ} |\phi_J\rangle\langle \phi_I|\ 
\end{align*}
onto the channel that corresponds to the
Coulomb branch parameter of the $SU(2)$ gauge
group. The bra and ket state of that projector
are inserted in the respective building blocks.
The problem thus reduces to computing various three
point functions
\begin{align*} 
\langle V(\phi^1_{I_1},z_1) V(\phi^2_{I_2},z_2)
V(\phi^3_{I_3},z_3)\rangle\ ,
\end{align*}
of primary or descendant fields. As pointed out
above, the subtlety lies in the choice of insertion
the points $z_i$. For linear and cyclic quivers, 
all the building blocks have only one or two
descendant fields inserted. Using the usual
coordinates on the sphere or torus, the descendant
fields are always inserted at 0 or $\infty$, that
is as bra and ket states,
and there is always a primary field inserted at 1.
Using this prescription the conformal block
agrees with the $SU(2)$ instanton partition function.

For trifundamental hypers the situation is more subtle.
We can insert three descendant fields at the points
$0,1,\infty$, but in general the result will not
agree with the instanton computation, since we
are using a different parametrization of the 
moduli space. Once expressed in IR variables,
the results will agree. To put it another way,
there will be a map between the moduli space coordinates
and the microscopic gauge coupling that will
make them agree. Or more geometrically, the CFT correlators define a
unique object on the Seiberg-Witten curve, that is independent on the
chosen parametrization of the complex structure moduli space of the
Gaiotto curve. 

The situation is much simpler for asymptotically
free gauge groups. In this case the conformal block
will agree with the instanton partition function immediately,
and will be essentially independent of the choice
of insertion points.

\subsubsection*{Comparison with Nekrasov partition function: one-loop
  factor}

As we consider theories with $\CN=2$ supersymmetry,
the full Nekrasov partition function has tree-level, 
one-loop and instanton contributions,
\be
 Z_{\rm Nek} = Z_{\textrm{clas}}(\tau_{\rm UV})  \,
 Z_{\textrm{1-loop}} \, Z_{\textrm{inst}}(\tau_{\rm UV}) \,.
\ee
The 4d/2d correspondence relates the purely representation-dependent
piece of the Liouville correlator
on the Gaiotto curve, that is the the conformal block, to the instanton
partition function of the corresponding gauge theory in the 
Omega-background. Adding the classical contributions to
the instanton partition function is crucial for finding good
properties under coordinate changes on the complex structure moduli
space of the Gaiotto curve (we spell this out explicitly in
section~\ref{sec:examples}). 
The one-loop factor can be
identified with the three point function of Liouville
theory. More properly, the full conformal block on the Gaiotto curve
should be identified with the Nekrasov partition function on $S^4$
\cite{AGT}. 

Let us check that this agreement continues to hold for Sicilian
quivers. The one-loop factor can be found as   
 a four-dimensional boson-fermion determinant in the
Omega-background. Equivalently, it may be obtained from the equivariant index
of the Dirac operator in the instanton background (see appendix~\ref{app:contourintegrands}). The resulting 
contribution for the (full)
$SU(2)$ trifundamental hypermultiplet is
\be \label{eqn:onelooptrif}
 Z_{\textrm{1-loop}}^{\rm 2trif} &=& \prod_{n, m = 1}^\infty \prod_{i, j, k=1}^2 \left(a_i + b_j +  c_k + \frac{Q}{2} + n \e_1 + m \e_2 \right)^{-1} \nn \\
 &\propto& \prod_{i, j,k=1}^2 \G_2 \left( a_i + b_j + c_k + \frac{Q}{2} \Big{|} \e_1, \e_2 \right),
\ee
where we take the Coulomb branch parameters $a_i=\pm a$, $b_j=\pm
b$ and $c_k=\pm c$ of the three $SU(2)$ gauge groups and $Q = \e_1 +
\e_2$. The Barnes' double gamma function $\G_2(x | \e_1, \e_2)$
regularizes the infinite product. 
The one-loop partition function for the $SU(2)$ trifundamental
half-hypermultiplet is given by a square-root of the
above expression. 

Agreement with the three-point function
of Liouville theory follows by the same argument
as for linear quivers \cite{AGT}. Namely,
the numerator of the DOZZ formula for the
Liouville three-point function contains the product
\be \label{eq:DOZZden}
\prod_{i, j,k=1}^2 \G_2 \left( a_i + b_j + c_k + Q/2 \right),
\ee
which equals the double trifundamental contribution in
equation~(\ref{eqn:onelooptrif}). Remember that the
product~(\ref{eq:DOZZden}) corresponds to 
the one-loop contribution of the Nekrasov partition function on
$S^4$, which splits into a chiral and anti-chiral contribution on
$\IR^4$.  Indeed, it is equal to the absolute value squared of the 
one-loop contribution for the $SU(2)$ trifundamental half-hyper,
which for example can be written as  
\be 
Z_{\textrm{1-loop}}^{\rm trif} = \G_2 (a+b+c+Q/2) \G_2(a + b-c + Q/2) \G_2 (a - b + c + Q/2) \G_2 (-a+b+c+Q/2).  \nn
\ee

\section{Examples}\label{sec:examples}

\begin{figure}[htbp]
\begin{center}
\includegraphics[width=3.7in]{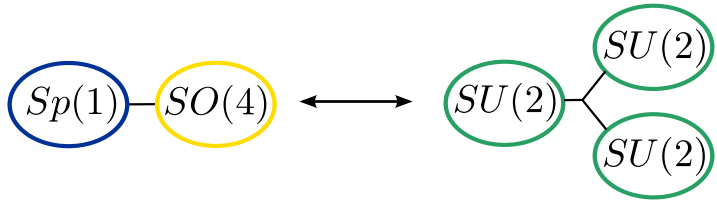}
\caption{From a gauge theory perspective the $Sp(1)-SO(4)$
  bifundamental, which is illustrated on the left, is equivalent to the $SU(2)$
  trifundamental, which is illustrated on the right, once we identify
  two of the $SU(2)$ gauge couplings. }\label{Fig:trifSOSpquiver} 
\end{center}
\end{figure}

\begin{figure}[htbp]
\begin{center}
\includegraphics[width=3.2in]{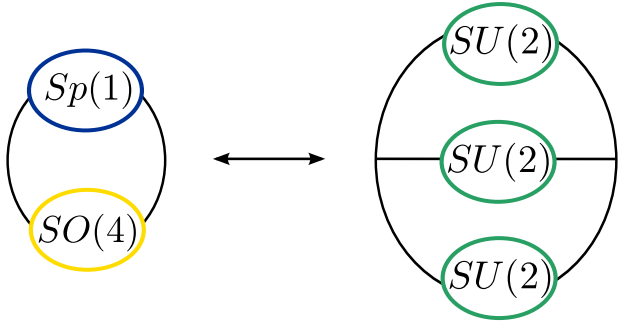}
\caption{From a gauge theory perspective the cyclic $Sp(1)-SO(4)$
  quiver, which is illustrated on the left, is equivalent to the genus
  2 $SU(2)$ quiver, which is illustrated on the right, once we identify
  two of the $SU(2)$ gauge couplings. }\label{Fig:doubletrifSOSpquiver} 
\end{center}
\end{figure}

In this section we test our proposal for extending the 4d/2d AGT
correspondence to Sicilian quivers in the two examples 
illustrated in
figure~\ref{Fig:trifSOSpquiver} and figure~\ref{Fig:doubletrifSOSpquiver}.


The quiver on the left in figure~\ref{Fig:trifSOSpquiver} consists of a single
$Sp(1)$ gauge group and a single 
$SO(4)$ gauge group coupled by a bifundamental $Sp(1)-SO(4)$
half-hypermultiplet. It is equivalent to an $SU(2)$ Sicilian quiver gauge theory
consisting of three $SU(2)$ gauge groups coupled by an $SU(2)$ trifundamental
half-hypermultiplet, illustrated on the right in figure~\ref{Fig:trifSOSpquiver}. 
The gauge couplings of both quivers are asymptotically free,
so that the instanton
partition function should agree directly with the CFT block
(\ref{CFTtrifund}) without any subtleties involving
a choice of coordinates. We check that this is indeed
the case up to order 3.

The quiver in figure~\ref{Fig:doubletrifSOSpquiver}
is a conformal
$Sp(1)-SO(4)$ gauge theory with two bifundamental $Sp(1)-SO(4)$
half-hypermultiplets, which is equivalent to a conformal $SU(2)$ Sicilian quiver
gauge theory
with three $SU(2)$ gauge groups coupled by two trifundamental $SU(2)$
half-hypermultiplets. 
Since the gauge theory is conformal, the results will depend on the choice 
of complex structure on the Gaiotto curve, and on the
instanton counting scheme. Our proposal tells us
which CFT configuration to choose to give direct agreement
with the $Sp(1)-SO(4)$ instanton partition function, and we check that
this indeed works up to order 3. 

The conformal $SU(2)$ gauge theory can
alternatively be described in terms of a massless full $SU(2)$ trifundamental
hyper. So we can find its instanton
partition function as well using the more conventional $U(2)$ instanton
counting scheme.\,\footnote{Notice that when we turn on the mass of this
  hypermultiplet, the theory does not have a string embedding
  anymore. This implies that we cannot find a Gaiotto curve. The
  Seiberg-Witten curve does exist, nevertheless, and can for example be found 
  through a semi-classical approximation of the instanton partition function.
  See \cite{Cecotti:2011rv} for a related discussion.
  }  We check that
if we use the $U(2)$ trifundamental instanton
counting scheme or choose different coordinates
in the conformal block the results do agree in
the IR. This confirms the general philosophy 
outlined above.

On a more technical level, 
the instanton counting formulae for the $Sp(1)-SO(4)$ quiver gauge
theories can be found in appendix~\ref{app:contourintegrands}. 
They are given by a multiple contour integral of
a meromorphic integrand.
This integrand consists of building blocks, each piece
coming from a component of the quiver gauge theory.
The contributions for the gauge theory nodes were already found in
\cite{NekrasovShadchin,Shadchin:2005mx}. We find the 
contribution for the
$Sp(1)-SO(4)$ bifundamental half-hypermultiplet as outlined in
section~\ref{sec:gauge}. We also make a proposal the integrand
for the full $SU(2)$ trifundamental hypermultiplet. 
To actually evaluate these contour integrands, that is
to find which of the poles contribute and to compute their residues, 
is an elaborate process, which we will describe later on.

\subsection{The $SU(2)$ trifundamental as a $Sp(1)-SO(4)$ bifundamental}
\begin{figure}[htbp]
\begin{center}
\includegraphics[width=2.in]{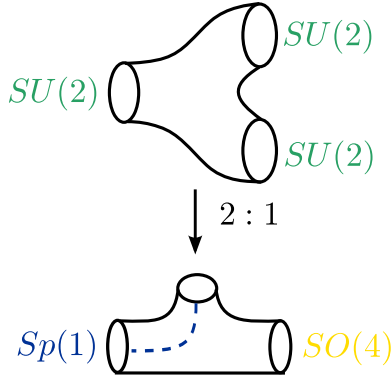} 
\caption{The Gaiotto curve for the $SU(2)$ trifundamental is a double
  covering of the Gaiotto curve for the $Sp(1)-SO(4)$
  bifundamental. (The corresponding quiver diagrams are illustrated in
  figure~\protect\ref{Fig:trifSOSpquiver}.) } \label{Fig:trifSOSpcurve} 
\end{center}
\end{figure}

In this section we compute the instanton partition function of the $Sp(1)-SO(4)$ quiver gauge theory with a single bifundamental half-hypermultiplet. The quiver diagram is given in figure~\ref{Fig:trifSOSpquiver} and the corresponding Gaiotto curve is illustrated in figure~\ref{Fig:trifSOSpcurve}. 

\subsubsection*{Computing the instanton partition function}
The instanton partition function of this theory is given by
\be
 Z_{\textrm{inst}}(q_1,q_2) = \sum_{k_1, k_2} q_1^{k_1} q_2^{k_2} \, Z_{k_1, k_2} 
\ee
with
\be \label{eq:sospquiver}
 Z_{k_1, k_2} = \oint \prod_{i = 1}^{n_1} d\phi_i \prod_{j=1}^{k_2}
 d\psi_j \, z^{Sp(1)}_{\textrm{vec}, k_1} (\phi) \,
 z^{SO(4)}_{\textrm{vec}, k_2} (\psi) \, z^{Sp(1)-SO(4)}_{\textrm{bifund}, k_1, k_2}( \phi, \psi) . 
\ee
Here $q_1$ and $q_2$ correspond to the exponentiated gauge couplings of the $Sp(1)$ and $SO(4)$ gauge group, respectively, and $k_1 = 2n_1 + \chi_1$. 
As mentioned above, the main problem is to find the correct prescription for the 
contour integral, and to evaluate the residues of the poles in question.
In the case of ordinary $SU(N)$ quiver gauge theories, 
the poles of the integrand only come from the vector multiplet 
contribution, and can be labeled by colored Young diagrams. 
In the case at hand, however, the $Sp(1)-SO(4)$ bifundamental 
does introduce additional poles, so that evaluating 
the contour integral becomes much more complicated. 
More precisely, besides the poles coming
from the $Sp(1)$ and the $SO(4)$ vector
multiplet, there are also poles
\be
 \psi_j &=& \pm \e_+~~\textrm{(when $k_2$ is
   odd)} \label{eq:bifundpole1}  \\
 \phi_i &=& \pm \psi_j \pm \e_+ \label{eq:bifundpole2} 
\ee
from the $Sp(1)-SO(4)$ bifundamental.
Note that these poles intertwine $Sp(1)$ poles and $SO(4)$ poles. 

A priori the integrals in (\ref{eq:sospquiver}) are over 
the real axis. We need to make a choice in moving
the poles away from the real axis and the closing
the contour.
The usual prescription is to move $\e_{1,2} \mapsto \e_{1,2} + i 0$ 
and then close the contour in the upper-half plane. 
We use this convention to deal with the vector multiplet poles. 
For the bifundamental poles, however,
we need to choose the opposite prescription $\e_+ \mapsto \e_+ - i0$. This
recipe originates from the description of the poles for
the massive full
bifundamental hypermultiplet. Similarly to  the pole prescription in
the $\CN=4$ ADHM construction, we introduce two additional equivariant
parameters $\e_3 = -\m - \e_+$ and $ \e_4 = \m - \e_+$, which we assume to
have positive imaginary parts.\,\footnote{See for example
  \cite{Bruzzo:2002xf} for a detailed discussion of the $\CN=4$ ADHM
  construction.} To find the pole prescription for the bifundamental
half-hypermultiplet, we just set 
the mass $\mu$ to zero (which identifies $\e_3=\e_4$).
We furthermore encounter poles of the form $ n \e_1 - m \e_3$ with $n \in
\half \IN$, $m \in \IN$, which we also need to include. Our prescription is
to take $\textrm{Im} (\e_\a - \e_\b) \gg 0$ if $\a > \b$ as in the
reference \cite{Moore:1998et}.

With this recipe we are set to evaluate the integral as the sum of
pole residues. For each integration variable $\phi_i$ or $\psi_j$ we
have a precise prescription, so that we can proceed integral by
integral.   
In practice it is useful to replace $\e_+ \mapsto -\e_3$ in the equations
\eqref{eq:bifundpole1} and \eqref{eq:bifundpole2} to avoid any source
of confusion.  After identifying the additional poles coming from the
bifundamental, we substitute back $\e_3 \mapsto -\e_+$ to evaluate the
integral.  
Note that the unrefined partition function can only be obtained by setting
$\e_1 + \e_2 = 0$ after we have performed the integration.

For the quiver gauge theory with a single $Sp(1)-SO(4)$ bifundamental
half-hyper (see figure~\ref{Fig:trifSOSpquiver}) the additional
bifundamental poles start to contribute at instanton number $k=(k_1, k_2)=(2,
1)$. There are 12 new poles at this order.  
To get agreement with the conformal block~(\ref{CFTtrifund}) it is
essential to include 
these extra poles.  
Interestingly, in the unrefined limit their contribution
happens to vanish, so that instanton counting becomes much simpler. 

\subsubsection*{Comparison with the three-point
  function~(\ref{CFTtrifund})} 
We identify the parameters of the conformal field theory and gauge theory to be
\be \label{eq:CFTtoSpSO}
 \Lambda_1 &=&  -\frac{q_1}{\e_1 \e_2}, ~ \Lambda_2 = -\frac{q_2}{16 \e_1 \e_2}, ~\Lambda_3 = \frac{q_2}{16 \e_1 \e_2} \nn \\
 h_1 &=& \frac{1}{\e_1 \e_2} \left( \frac{Q^2}{4} - a^2 \right) ,  \\
 h_2 &=& \frac{1}{\e_1 \e_2} \left(\frac{Q^2}{4} - \left( \frac{b_1 + b_2}{2} \right)^2 \right) , \nn \\
 h_3 &=& \frac{1}{\e_1 \e_2} \left(\frac{Q^2}{4} - \left( \frac{b_1 - b_2}{2} \right)^2 \right) \nn, \\
 c &=& 1 + \frac{6 (\e_1 + \e_2)^2}{\e_1 \e_2} \nn ,
\ee
where $c$ is the central charge and $h_i$ are the conformal weights of the vertex operators. 
Comparing this with the three-point function~(\ref{CFTtrifund}) we find
\be\boxed{
Z_{\textrm{inst}}(q_1,q_2)= Z_{\textrm{CFT}}(q_1,q_2) Z_{\textrm{spur}}^{(1)}\,, }
\ee
up to $k=(k_1, k_2) = (2, 2)$,
with the spurious factor
\be
 Z^{(1)}_{\textrm{spur}} = \exp\left( \frac{q_1}{2 \e_1 \e_2} \right)
 \exp \left(\frac{q_2}{8\e_1 \e_2} \right)\,. \notag
\ee

\subsection{Genus two quiver through $Sp(1)-SO(4)$ instanton counting} 
 
\begin{figure}[htbp]
\begin{center}
\includegraphics[width=1.5in]{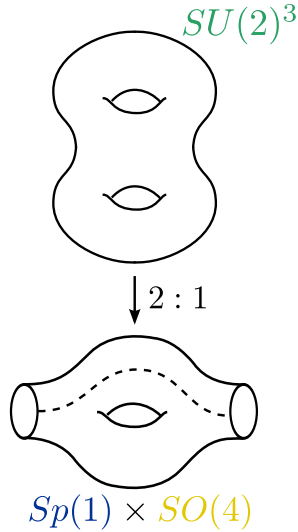}
\caption{The Gaiotto curve for the $SU(2)$ genus 2 quiver is a double
  covering of the Gaiotto curve for the $Sp(1)-SO(4)$
  cyclic quiver.  (The quiver diagrams are illustrated in
  figure~~\protect\ref{Fig:doubletrifSOSpquiver}.) }\label{Fig:doubletrifSOSpcurve} 
\end{center}
\end{figure}

The Sicilian quiver theory for genus two Gaiotto curve is an $SU(2)^3$
theory with two trifundamental half-hypers. We can also view it as a
$Sp(1)-SO(4)$ theory with two bifundamental half-hypers (see figure
\ref{Fig:doubletrifSOSpquiver}). As illustrated in figure
\ref{Fig:doubletrifSOSpcurve}, the corresponding Gaiotto curve is a
torus with two punctures.  
The genus two curve is the double cover of the genus one curve with
two branch points.  

The partition functions will depend on the choice of coordinates of the Riemann
surface, because the gauge theory is conformal. 
Using our proposal for the $Sp(1)-SO(4)$ AGT correspondence, we can
find a choice of coordinates on the complex moduli space of the genus
two Gaiotto curve that matches the instanton partition 
function with the conformal block directly, without a UV-UV map. 

\subsubsection*{Computing the instanton partition function}

Since a single full hypermultiplet can be obtained by combining
two half-hypermultiplets, there are two different ways to compute the instanton
partition function: either as a single full bifundamental hyper  or as
two bifundamental half-hypers.  
Using the first method we start from the massive full hyper.  
The term at order $(k_1, k_2)$ is given by
\be
 Z^{(1)}_{k_1, k_2} = \oint \prod_{i=1}^{n_1} d\phi_i
 \prod_{j=1}^{k_2} d\psi_j \, z^{Sp(1)}_{\textrm{vec}, k_1} (\phi, a) \,
 z^{SO(4)}_{\textrm{vec}, k_2} (\psi, b) \, z^{Sp(1), SO(4)}_{\textrm{2bif} , k_1, k_2} (\phi, \psi, \m) . 
\ee
where the explicit form of the integrand is given in
appendix~\ref{app:contourintegrands}.  

We can evaluate the contour integral using the two equivariant
parameters, $\e_3 = -\m - \e_+$ and $\e_4 = \m - \e_+$. The massive
full bifundamental introduces the additional poles
\be
 \psi_j &=& \pm \e_3, \pm \e_4~~\textrm{(when $k_2$ is odd)} \label{eq:dbifundpole1}  \\
 \phi_i &=& \pm \psi_j \pm \e_3 \label{eq:dbifundpole2}  \\
 \phi_i &=& \pm \psi_j \pm \e_4 \label{eq:dbifundpole3} .
\ee
Compared to the single bifundamental half-hyper in the previous example the
additional parameter $\e_4$ introduces extra poles. For example, there
are in total 28 new poles with non-vanishing residues at $k=(1, 2)$.  
Again the contribution from these new poles happens to vanish in the
unrefined limit, whereas it is crucial to include them in the refined
setup.  

Using the second method we start with the contour integral
corresponding to two massless bifundamental
half-hypers
\be
 Z^{(2)}_{k_1, k_2} = \oint \prod_{i=1}^{n_1} d\phi_i
 \prod_{j=1}^{k_2} d\psi_j \, z^{Sp(1)}_{\textrm{vec}, k_1} (\phi, a)
 \, z^{SO(4)}_{\textrm{vec}, k_2} (\psi, b) \left( z^{Sp(1), SO(4)}_{\textrm{bif} , k_1, k_2} (\phi, \psi) \right)^2,  
\ee
where the explicit form of each of the building blocks is given in
appendix~\ref{app:contourintegrands}. The poles of this integral are
simpler to enumerate, since there is just one new equivariant
parameter $\e_3 = -\e_+$. There are 10 new poles with non-vanishing
residues at $k=(1, 2)$. As they should, both computations indeed give
the same result once we set the mass $\m = 0$.

\subsubsection*{CFT computation}

Let us now compute the conformal block for the genus two surface.
The most straightforward guess is to imitate
(\ref{CFT6pct}) by taking
\be \label{CFTg2naive}
Z = \sum_{I_1,I_2,I_3}\sum_{J_1,J_2,J_3} 
K^{-1}_{I_1J_1}K^{-1}_{I_2J_2}K^{-1}_{I_3J_3}\,
\langle \phi^{a_1}_{I_1}|V(\phi^{a_2}_{I_2},1)
|\phi^{a_3}_{I_3}\rangle
\langle \phi^{a_1}_{J_1}|V(\phi^{a_2}_{J_2},1)
|\phi^{a_3}_{J_3}\rangle\, q_1^{I_1}q_2^{I_2}q_3^{I_3} .\
\ee
By the general remarks above, this expression 
corresponds to a particular parametrization of the
moduli space of the genus two surface. Presumably
there should be a corresponding regularization
scheme on the gauge theory side. In particular, the conformal block
(\ref{CFTg2naive}) should agree with any instanton
computation in the IR. 

We want to do a bit better than that however:
we want to find an expression which agrees with
our instanton computation of the genus two surface as 
a cyclic $Sp(1)-SO(4)$ quiver in the UV.
The AGT correspondence for $Sp/SO$ quivers was
worked out in \cite{HKS}. Let us briefly
summarize the relevant facts.

The Gaiotto curve of the cyclic $Sp/SO$ theory
is a torus with a $\IZ_2$ branch cut running
between two branch points. The double cover
of this curve is the genus two curve
where two of the moduli are equal, see
figure~\ref{Fig:doubletrifSOSpcurve}.  
The $\CW$-algebra of the theory is a double copy
of the Virasoro algebra, where the $\IZ_2$-twist
exchanges the two copies.
The conformal block of this configuration on 
the torus with total modulus $q_1^2 q_2$ is given by
\be\label{CFTSpSOcyclic}
\Tr\left[ \sigma(1)P_{a_1}\sigma(q_1^2)P_{a_2,a_3} (q_1^2q_2)^{L_0}\right]\,.
\ee
Here $\sigma$ is the $\IZ_2$-twist vacuum, and we take the
branch cut to go from $\sigma(1)$ through $P_{a_1}$ to $\sigma(q_1^2)$.
$P_{a_1}$ is the projector onto the twisted representation
coming from the primary field $\phi^{1}$. As the primary field
$\phi^{1}$ transforms in a twisted
representation, it is indeed characterized by a single
parameter $a_1$. On the other hand, $P_{a_2,a_3}$ is
the projector onto the untwisted representation characterized
by two parameters $a_2$ and $a_3$.

\begin{figure}[htbp]
\begin{center}
\includegraphics[width=6in]{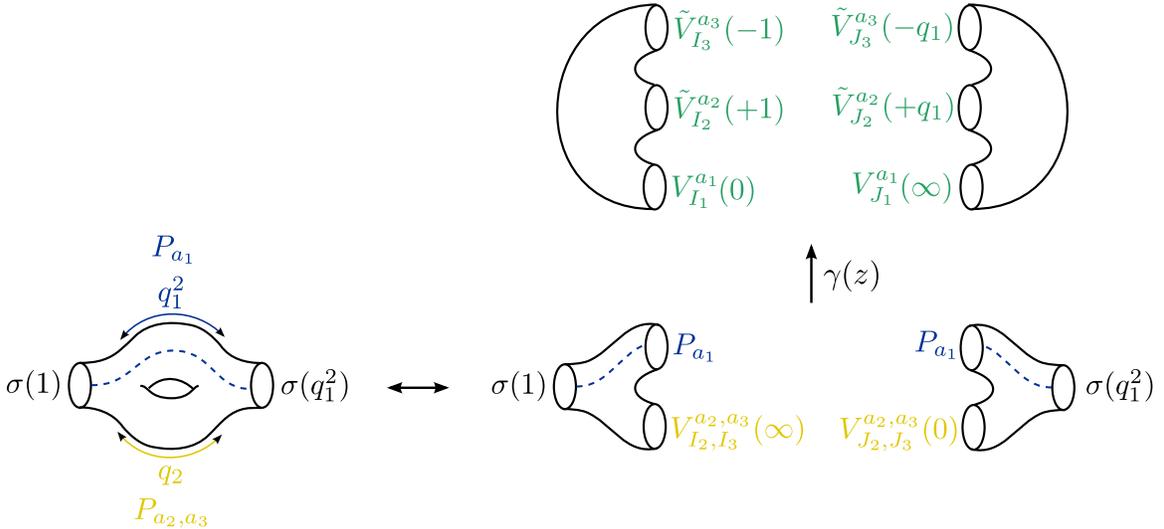}
\caption{The map $\gamma(z)$ relates the $\CW$-block on the twice
  punctured torus (which computes the double $Sp(1)-SO(4)$ instanton
  partition function) to a Virasoro block on its double cover, a genus
two curve. More precisely, we first cut open the torus along the
$SO(4)$ tube and insert a complete basis of states
$V_{I_2,I_3}^{a_2,a_3}$ in the untwisted representation labeled by
$a_2$ and $a_3$. Then we map this onto
a genus two surface using the map $\gamma$, and  insert a
complete basis of states $V^{a_1}_{I_1}$ in the Virasoro
representation labeled by $a_1$. 
%Note that we need to apply an additional transformation to map
%this configuration to the standard Virasoro building blocks (which
%compute the $SU(2)$ trifundamental instanton partition function).
}\label{Fig:doubletrifCFT} 
\end{center}
\end{figure}

Computing correlators of twisted representations can be done
by going to the cover of the surface. Here, the $\CW$-algebra on
the cover is a single copy of the Virasoro algebra, and
the problem reduces to the computation of the standard
conformal block on the genus two surface.
Although conceptually straightforward, this procedure
leads to some technical subtleties. The main problem is
that we apply the cover map to compute 
correlators with Virasoro descendants,
which leads to correction terms. Let us therefore
spell out precisely the cover map and the ensuing
correlation functions.

To map the four point function (\ref{CFTSpSOcyclic})
to the cover, we use the map $\gamma$
\be \label{CFTcovermap}
z \mapsto \gamma(z) = \pm \sqrt{\frac{z-q_1^2}{z-1}}\ ,
\ee
where the sign determines the branch of the cover.
This map indeed has branch points at $z=1$ and $z=q_1^2$,
and it maps the operators at 0 and $\infty$ to 
$\pm q_1$ and $\pm 1$. This is illustrated in
figure~\ref{Fig:doubletrifCFT}. In a second step, we want
to reduce everything to standard building blocks, 
that is three-point functions on the sphere
with operators inserted at $1,-1$ and $0$.

\begin{table}
\be \label{mapdiagram}
\begin{array}{ccccc}
\multicolumn{5}{c}{\langle V^{a_2,a_3}_{I_2,I_3}(\infty) \sigma(1) P_{a_1} \sigma(q_1^2) 
V^{a_2,a_3}_{J_2,J_3}(0)\rangle} \\
\\
&&\downarrow&\multicolumn{2}{l}{z\mapsto \gamma(z)} \\
\\
\multicolumn{5}{c}{
\langle \tilde V^{a_2}_{I_2}(1)\tilde V^{a_3}_{I_3}(-1)|V^{a_1}_{I_1}\rangle
\langle V^{a_1}_{J_1}|\tilde V^{a_2}_{J_2}(q_1)\tilde V^{a_3}_{J_3}(-q_1)\rangle }\\
\\
\swarrow&&&&\searrow z\mapsto q_1/z\\
\\
\langle \tilde V^{a_2}_{I_2}(1)\tilde V^{a_3}_{I_3}(-1)|V^{a_1}_{I_1}\rangle&\qquad &\ &\qquad &
\langle \tilde V^{a_2}_{J_2}(1)\tilde V^{a_3}_{J_3}(-1)|V^{a_1}_{J_1}\rangle
\end{array}\nn
\ee
\caption{Sequence of maps used for computing the
cyclic $Sp(1)-SO(4)$ quiver.\label{covermaps}}.
\end{table}

The total sequence of maps is shown in table~\ref{covermaps}.\,\footnote{The $\tilde{V}(\pm q_1)$'s in
  the second line of table~\ref{covermaps} are slightly different from
  the $\tilde{V}(\pm 1)$'s, which are  
  defined in equation~(\ref{eqn:newvertexop}).}
Introducing the notation
\be
\mathbf{C}^{a_2,a_3;a_1}_{I_2,I_3,I_1}(q_1)=
\langle \tilde V^{a_2}_{J_2}(1)\tilde
V^{a_3}_{J_3}(-1)|V^{a_1}_{J_1}\rangle\,
q_1^{2(h_{a_2}+h_{a_3}+I_2+I_3)} , \notag
\ee
the conformal block is given by
\be \label{eq:CFTg2}
Z_{\textrm{CFT}} = \sum_{I_i,J_i} K^{-1}_{I_1J_1}K^{-1}_{I_2J_2}K^{-1}_{I_3J_3}
\mathbf{C}^{a_2,a_3;a_1}_{I_2,I_3,I_1}(q_1)\mathbf{C}^{a_2,a_3;a_1}_{J_2,J_3,J_1}(q_1)
 q_1^{I_1} q_2^{I_2+I_3} .
\ee
Note that the three-point functions $\mathbf{C}(q_1)$ depends on $q_1$ in a 
non-trivial way. 
They are obtained by
acting with $q_1$-dependent coordinate transformations,
so that the vertex operators $\tilde V$ depend on $q_1$. 
Since (\ref{CFTcovermap}) is not a M\"obius transformation,
we need the following generalization \cite{Gaberdiel:1994fs}  
of (\ref{Mobius}): 
\be\label{CFTtrafo}
V(\phi,z) \mapsto V \left( \left[\prod_{n=1}^\infty
  \exp\left(\frac{T_n(z)}{f'(z)^n}L_n\right)\right] f'(z)^{L_0} \phi ,f(z) \right) ,
\ee
which holds for $z$ away from singular points.
Here we take all products to go from left to right.
The functions $T_n(z)$ are defined recursively. The first two
are given by
\be
T_1(z)=\frac{f''(z)}{2f'(z)}\ , \qquad T_2(z)=\frac{1}{3!}\left(\frac{f'''(z)}{f'(z)}
-\frac{3}{2}\left(\frac{f''(z)}{f'(z)}\right)^2\right)\,. \notag
\ee
For the transformations 
in (\ref{mapdiagram}) the new vertex operators are given by
\be\label{eqn:newvertexop}
\tilde V^{a}_{I}(\pm1)= \left(\pm\frac{1-q_1^2}{2q_1^2}\right)^{h_{a}+I}
V\left(\exp\left[\pm\frac{3+q_1^2}{2(1-q_1^2)}L_1\right] 
\exp\left[\frac{1}{4}L_2\right]\cdots \phi^a_I,\pm1 \right)\,, 
\ee 
where the dots signify exponential factors involving higher $L_n$.

When we compare the genus two conformal block \eqref{eq:CFTg2} with the instanton counting result, we 
find that they indeed agree up to order $k=(1, 2)$:
\be
\boxed{
Z_{\textrm{inst}}(q_1,q_2)=Z_{\textrm{CFT}}(q_1,q_2)Z_{\textrm{spur}}^{(2)}
\, ,
}
\ee 
where the spurious factor 
\be
Z_{\textrm{spur}}^{(2)} =  1-\frac{3 \left(\e_1+\e_2\right){}^2 }{8 \left(\epsilon _1 \epsilon _2\right)} q_2 +\frac{3 \left(\epsilon _1+\epsilon _2\right){}^2 \left(3 \epsilon _1^2-2 \epsilon _1 \epsilon _2+3 \epsilon _2^2\right) }{128 \epsilon _1^2 \epsilon _2^2} q_2^2 + \ldots\notag
\ee
 does not depend on physical parameters.

\subsection{Alternative prescriptions for the genus two quiver}

In the previous example we chose the coordinates
of the genus two Gaiotto curve in such a way that we obtained
direct agreement between the conformal block
and the $Sp(1)-SO(4)$ instanton counting.
If we choose different coordinates, or if we use
a different instanton counting scheme, then 
the result will be different. In the infrared, however,
all versions should agree. Put differently,
we should be able to find a map between the UV couplings that
make two results agree. Let us show 
that this philosophy is correct in two examples.

\subsubsection*{Comparison using a different conformal block}

First, we can use the ``naive'' conformal block \eqref{CFTg2naive} 
to compute the genus two correlator.\,\footnote{We additionally checked that
  genus two correlator is independent on the choice of internal
  punctures up to a UV-UV mapping.}  We indeed
find agreement between 
the $Sp(1)-SO(4)$ instanton partition function and this genus two
conformal block 
using the UV-UV map
\be \label{eq:g2UVUVmap}
\tilde{q}_1 &=& q_1 - \frac{1}{4} q_1 q_2 - \frac{1}{2} q_1^2 - \frac{1}{16} q_1 q_2^2 + \frac{1}{4} q_1^2 q_2 + \frac{3}{16} q_1^3 + O(q^4) , \nn \\
\tilde{q}_2 &=& \frac{1}{16} q_2 - \frac{1}{16} q_1 q_2 + \frac{3}{128} q_2^2 - \frac{1}{128} q_1 q_2^2 + \frac{3}{128} q_1^2 q_2 + \frac{55}{4096} q_2^3 + O(q^4) , \\
\tilde{q}_3 &=& \frac{1}{16} q_2 + \frac{1}{16} q_1 q_2 + \frac{3}{128} q_2^2 + \frac{1}{128} q_1 q_2^2 + \frac{3}{128} q_1^2 q_2 + \frac{55}{4096} q_2^3 + O(q^4) . \nn
\ee

To compare the instanton partition functions with different
parametrization of couplings, it is important to include tree-level
pieces.\footnote{One-loop factors are not relevant since they do not
involve the gauge couplings.} Similarly we also need to include
the tree-level piece of the conformal block. In the case at hand we have  
\be
 Z_{\textrm{tree}}^{Sp-SO}(q) = q_1^{- \frac{a^2}{ \e_1 \e_2} }
 \left(\frac{q_2}{16}\right)^{- \frac{1}{2 \e_1 \e_2}( b_1^2 + b_2^2
   )} \quad {\textrm{and}} \quad 
  Z_{\textrm{tree}}^{\textrm{CFT}}(\tilde{q}) = \tilde{q}_1^{-\frac{a_1^2}{\e_1\e_2}} \, 
  \tilde{q}_2^{-\frac{a_2^2}{\e_1\e_2}}  \, 
  \tilde{q}_3^{-\frac{a_3^2}{\e_1\e_2}}\,. \notag
\ee
Note that these two factors are related by 
the identifications (\ref{eq:CFTtoSpSO}) and
the mappings $a_2
\to \half (b_1 + b_2)$ and $ a_3 \to \half (b_1 - b_2)$. 
Using the UV-UV  mapping \eqref{eq:g2UVUVmap} and the above
identification of Coulomb parameters, we find that the ratio
of $Z_{\textrm{Nek}}$ and $Z_{\rm CFT}$ is given by the
spurious factor
\be
 Z_{\textrm{spur}}^{(3)} = 1 - \frac{(\e_1 + \e_2)^2}{8 \e_1 \e_2} q_1 - \frac{(\e_1 + \e_2)^2}{4 \e_1 \e_2} q_2 + \frac{(\e_1 + \e_2)^2 (\e_1^2 + 3\e_1 \e_2 + \e_2^2)}{32 \e_1^2 \e_2^2} q_1 q_2 + \cdots .\notag
\ee

\subsubsection*{Comparison using a different method of instanton counting}

As another test of this philosophy on the gauge theory side, 
we can compute
the instanton using the $U(2)$
instanton counting  scheme. This is possible since two $SU(2)$ trifundamental
half-hypers combine into a single massless $SU(2)$ full
trifundamental 
hyper.
Apart from a non-trivial UV-UV mapping, we expect the
$U(2)$ instanton and the CFT computation to differ by a non-trivial
$U(1)$ factor. This $U(1)$ factor should not depend on any of the
Coulomb parameters, after we enforce the tracelessness condition on all
three gauge groups.    

Since the trifundamental hyper is in the fundamental representation of
the three gauge groups
$SU(2)_A \otimes SU(2)_B \otimes SU(2)_C$, we find the contour
integrand by considering the tensor product $$\CE_A \otimes \CE_B
\otimes \CE_C \otimes \CL$$ of the three $U(2)$
universal bundles $\CE_{A/B/C}$ over the product of ADHM moduli spaces
$\CM_{U(2)_A} \times
\CM_{U(2)_B} \times \CM_{U(2)_C} \times \IR^4$ with
the half-canonical bundle $\CL$ over $\IR^4$.
Recall that the restriction of each universal bundle to a self-dual
connection $\CA$ is just the corresponding
instanton bundle $\CE|_{\CA} = E$ over $\IR^4$ (see for example
\cite{HKS} for more 
details).  The equivariant weights contributing to the Euler class
$e_{\mathbf{T}}(\CV)$ of the bundle $\CV$ of Dirac zero modes can then be
found from the equivariant Dirac index
\begin{align}
\mbox{Ind}_{\mathbf{T}}(\slashed D)^{g=2}_{U(2)^3} = \int_{\IC^2}
\Ch_{\mathbf{T}} (\CE_A \otimes \CE_B \otimes \CE_C \otimes \CL) \,
\Td_{\mathbf{T}}(\IC^2). 
\end{align} 
The resulting contour integrand
can be found in appendix~\ref{app:contourintegrands}.
For up to two instantons, it reproduces the partition
functions for the bifundamental.
For three non-zero instanton numbers the evaluation 
of the contour integral becomes tricky, 
because, unlike for $U(2)$ bifundamentals, many additional
poles appear. It would be interesting to find an elegant
prescription for the additional poles that yields agreement with the
CFT. 

We can still compare the instanton partition function up to second
order. We find that up to this order it agrees 
with the conformal block~(\ref{CFTg2naive}),
when we use the map
\be
\tilde{q}_{A} &=& q_A + 2 q_A^2 q_B + 6 q_A q_B q_C + \CO(q^4) \notag\\
\tilde{q}_{B} &=& q_B + 2 q_A q_B + 2 q_B q_C + 8 q_A q_B q_C + 3 q_A^2
q_B - 2 q_A q_B^2 + 2 q_B^2 q_C + 3 q_B q_C^2 + \CO(q^4) \notag \\
\tilde{q}_{C} &=& q_C + 2 q_B q_C + 6 q_A q_B q_C + 3 q_B^2 q_C + 2
q_B  q_C^2 + \CO(q^4) \notag
\ee
between gauge coupling constants $\tilde{q}_{A/B/C}$ and complex
structure parameters $q_{A/B/C}$, and up to a (unrefined) $U(1)$ factor
\be 
Z_{\rm U(1)} = 1 + q_A q_B + q_B q_C + q_C q_A + \cdots 
\ee
Again, we need to include the classical contributions here.
The non-trivial UV-UV mapping is expected since the
conformal three-point function only reduces directly to a bifundamental
contribution when two of its punctures are set at the positions $0$
and $\infty$, and the primary vertex operator is inserted at $1$.

\section{Discussion}\label{sec:discussion}

Let us briefly summarize our results and discuss 
some open questions. In this paper we  
extended the AGT correspondence
to Sicilian quivers. To do this we first pointed out that
the instanton 
partition function corresponding to the $SU(2)$ trifundamental 
cannot be found using the more conventional
$U(2)$ instanton counting scheme, but
should be computed using either an $Sp(1)^3$ or an $Sp(1)-SO(4)$ instanton
counting scheme. 
We also argued that since the former does not have a string
theory embedding, it is more natural to use the latter
to find a corresponding CFT configuration.

As one of our main result we found that the instanton partition function for
the $SU(2)$ trifundamental interaction is equal to the three-point
function of Gaiotto states
\begin{equation}\label{eqn:CFTtrif}
\boxed{\langle h_1, \Lambda_1 | V( | h_2, \Lambda_2 \rangle, 1) |  h_3,
\Lambda_3 \rangle.} \ 
\end{equation} 
We have verified that it satisfies various
consistency checks, such as symmetry under exchange
of the gauge groups and reduction to
the $SU(2)$ bifundamental. 
Furthermore, it agrees with the instanton
partition function for the $Sp(1)-SO(4)$ bifundamental. 

We have also proposed a construction for the Virasoro conformal
block corresponding to any asymptotically free or conformal
generalized $SU(2)$ quiver gauge theory. 
This proposal again passes all consistency checks,
such as having a
a proper genus-expansion in the parameter $\hbar$, and
it has the expected properties in terms of UV vs. IR 
parameters. These
are non-trivial properties for the conformal blocks, and it would be
interesting to find a CFT explanation.  
Secondly, we have checked the proposal against $Sp(1)-SO(4)$
instanton counting in several Sicilian examples. To get an exact
agreement of the instanton partition functions with the conformal
blocks, we have used the string embedding of the $Sp(1)-SO(4)$ gauge
theories to find the correct parametrizations of the conformal blocks.   

There remain several interesting questions. It would for example be insightful to
find a (geometric, or gauge theoretic) explanation for the asymmetry
of the three-point function with 
three descendants. Relatedly, one can try to come up with a 
prescription for the parametrization of the conformal blocks dual to
conformal Sicilian quivers that
agrees on the nose with the instanton partition function. One can also
wonder whether it is possible to perform instanton counting for any
choice of coordinates on the complex structure moduli space. Why does
instanton counting choose the particular parametrization it chooses?  
It would furthermore be interesting to look for a CFT object on the
Seiberg-Witten curve that agrees with the IR partition function. 

From the instanton counting perspective it is also curious that every
instanton partition function can be decomposed into interactions with
three or fewer gauge groups. This translates to the statement that
any cohomology class on the instanton moduli space can be written as a
product of only a few elementary classes. It would be interesting to
understand this better, and necessary for a complete understanding of
instanton counting for Sicilian quiver theories. 

An exciting extension would be a verification of our proposal
through geometric engineering \cite{Benini:2009gi} and the (refined)
topological vertex \cite{Kozcaz:2010af}. It is not even obvious that the
toric diagrams corresponding to Sicilian quivers indeed have the
correct description under the decoupling of gauge groups. Furthermore,
in this setting gauge groups are engineered using combinations of
D4 and NS5-branes. This is very unusual from the perspective of geometric
engineering, and requires a better understanding.  Another fruitful
direction would be the inclusion of BPS operators in these Sicilian
gauge theories. Finally, a full
verification of our proposal requires a more detailed study of
instanton counting for the $Sp(1)^3$ 
trifundamental half-hypermultiplet
\cite{Nekrasov-Pestun}.\,\footnote{Starting from the $Sp(1)$ 
  instanton formalism of \cite{NekrasovShadchin,Shadchin:2005mx} and
  using the tools developed in this paper, we find an expression
  for the contour integral
  which has odd  properties. It is quite possible that these may be caused by 
  the lack of a string theory embedding of the $Sp(1)^3$ trifundamental
  half-hypermultiplet.}

\acknowledgments

We would like to thank Fernando Alday, Alexander Braverman, Andrew
Dancer, Nikita 
Nekrasov, Sara Pasquetti, Vasily Pestun, Yuji Tachikawa for enlightening discussions
and correspondence. 
LH would like to thank the support of Sergei Gukov and the
hospitality of the Mathematics Institute of the University of
Oxford. The work of LH is supported by a NWO Rubicon fellowship and by NSF
grant PHY-0757647. 
The work of CAK is supported by a John A.~McCone Postdoctoral
Fellowship. This work is in addition supported in part by the DOE grant
DE-FG03-92-ER40701. 
\\

%%%%%%%%%%%%%%%%%%%%%%%%%%%%%%%%%%%%%%%%%%%%%%
\vskip 1cm
\centerline{\Large\bf Appendix}
\appendix

\section{More about the trifundamental half-hypermultiplet}\label{app:flavorenh}
In this appendix we explicitly show the reduction of the Lagrangian
for the $SU(2)$ trifundamental half-hypermultiplet to bifundamental and
fundamental hypermultiplets when we Higgs one or more of the $SU(2)$
gauge groups. Before starting this argument, let us quickly remind
ourselves about flavor symmetry enhancement for (pseudo-)real
representations.      

\subsubsection*{Flavor symmetry enhancement}

Let us briefly explain a way to understand the enhancement of flavor
symmetries for matter transforming in a  (pseudo-)real
representation. Although this is not of direct importance for this
paper, it will be useful as background and in the following.  

First of all, recall the familiar statement of flavor symmetry
enhancement. The flavor symmetry group of $N$ hypers in a
real representation of the gauge group is enhanced from $U(1)^N$ to
$Sp(N)$, whereas the flavor symmetry of $N$ hypers in a pseudo-real
representation is enhanced to $SO(2N)$. 
The cases that are important for us is the
single hyper in the
bifundamental of  $SU(2)$ which has enhanced flavor symmetry $Sp(1)=SU(2)$,
and two hypers in the fundamental which enhance to $SO(4)=SU(2)\times SU(2)$.

A pseudoreal representation is characterized by an
antilinear map $\sigma_G$ such that $\sigma_G^2=-1$.
This map corresponds to the complex conjugation,
so that in case of the fundamental or adjoint 
representation $\sigma_G^{-1} T \sigma_G = T^* = -T^t$ for 
$T \in \mathfrak{g}$. For real representations
the only difference is that $\sigma_G^2=1$. Note
that $\sigma_G$ is automatically unitary.

The basic idea behind the enhancement is that the $N$-dimensional `flavor
vector' $Q_i$ is enlarged to a $2N$-dimensional vector $(Q_i, \sigma_G \tilde{Q}_i)$
(which still is in the representation $R$ of the gauge group).
What needs to be shown is that the terms in the Lagrangian
are invariants of $SO(2N)$ or $Sp(N)$.

For a single
hypermultiplet in the fundamental representation of $SU(2)$, which is
pseudo-real, 
the kinetic term in the Lagrangian $\CL_{\rm{fh}}$ can be rewritten as
\begin{align}\notag
Q^\dagger e^V Q + \tilde{Q}^t e^{-V} \tilde{Q}^* 
&= Q^\dagger e^V Q + \tilde{Q}^\dagger e^{-V^t} \tilde{Q} \\ 
&= Q^\dagger e^V Q + \tilde{Q}^\dagger \sigma_G^{-1} e^V \sigma_G
\tilde{Q}  \notag \\
&= \left( \begin{array}{cc} Q^{\dagger}, &  \tilde{Q}^\dagger \sigma_G^\dagger \end{array}
\right) e^V \left( \begin{array}{c} Q \\ 
    \sigma_G \tilde{Q} \end{array} \right)\ . \label{eqn:kin-pr-enh}
\end{align}
The Yukawa coupling in the Lagrangian $\CL_{\rm{fh}}$ is
proportional to  
\begin{align}\notag
 2 \, \tilde{Q}^t \Phi Q  &=  Q^t \Phi^t \tilde{Q} + \tilde{Q}^t \Phi Q  \\
&= Q^t \sigma_G \Phi \sigma_G \tilde{Q} + \tilde{Q}^t \Phi Q \notag \\
&= \left( \begin{array}{cc}  Q^t, & \tilde{Q}^t\sigma_G^t \end{array}
\right) \, \sigma_G \Phi \left( \begin{array}{cc} 0 & 1 \\ 1 & 
    0 \end{array} \right) \left( \begin{array}{c} Q \\ 
    \sigma_G \tilde{Q} \end{array} \right). \label{eqn:Yuk-pr-enh}
\end{align}
In both cases we used the fact  $\sigma_G^{-1} T \sigma_G = -T^t$ for 
$T \in \mathfrak{g}$.

To see that (\ref{eqn:Yuk-pr-enh}) is an $SO(2)$ invariant,
we can make a change of basis to $Q^{\pm}=Q\pm i\sigma_G \tilde{Q}$.
The enhanced flavor group $SO(2)$ then acts in the fundamental
on this new basis, and both (\ref{eqn:kin-pr-enh})
and (\ref{eqn:Yuk-pr-enh}) are the standard diagonal
invariants.
This argument generalizes in a straightforward way
to an arbitrary number of hypers $Q_i$.

A similar argument works for real representations,
such as for the bifundamental in $SU(2)_A \times SU(2)_B$.
In this case the kinetic term can written exactly
in the form (\ref{eqn:kin-pr-enh}) as well. The Yukawa term picks up
a minus sign, due to the fact that now $\sigma_G^2=1$,
so that the invariant is found by replacing
$$\left(\begin{array}{cc} 0 & 1 \\ 1 & 0 \end{array}\right)
\mapsto \left(\begin{array}{cc} 0 & \one_N \\ -\one_N & 0 \end{array}\right)$$
in (\ref{eqn:Yuk-pr-enh}). The Lagrangian is thus indeed invariant under
$Sp(N)$. (Note that by $Sp(N)$ we actually 
mean $USp(N)=U(2N)\cap Sp(2N,\IC)$ here.)

\subsubsection*{Reduction of the $SU(2)$ trifundamental
  half-hypermultiplet} 

We first show that the Lagrangian for the $SU(2)$ trifundamental
half-hyper reduces to that of a massive $SU(2)$ bifundamental hyper
when one of the three $SU(2)$ gauge groups is Higgsed.  
Start with the Yukawa terms 
\be
W = \e^{bb'} \e^{cc'} Q_{abc} \Phi_1^{aa'} Q_{a'b'c'}  + \e^{aa'} \e^{cc'} Q_{abc} \Phi_2^{bb'} Q_{a'b'c'} + \e^{aa'} \e^{bb'} Q_{abc} \Phi_3^{cc'} Q_{a'b'c'}   
\ee
in the Lagrangian of the $SU(2)$ trifundamental half-hyper (as derived
in section~\ref{sec:gauge}). The first gauge group is Higgsed by setting 
$(\Phi_1)^{a}_{~a'} = m_1 (\sigma_3)^{a}_{~a'}$, so that the
superpotential $W$ reduces to
\begin{align}\label{eqn:HiggsYuktrif}
W &= m_1 (\sigma_3)^{a'}_{~a} Q^{abc}  
Q_{a'b c}  - \e^{c c'}  Q^a_{~bc} \Phi_2^{bb'} Q_{ab'c'} -
\e^{b b'}  Q^a_{~bc} \Phi_3^{cc'} Q_{a b' c'}. 
\end{align}
When identifying
\be\label{eqn:trifbifid}
 Q_{bc}  &\equiv& Q_{1bc} = -Q^{2}_{~bc}    \\
\tilde{Q}_{bc} & \equiv&  Q_{2bc} = Q^1_{~bc}  \notag 
\ee
in equation~(\ref{eqn:HiggsYuktrif}), we indeed recover the Yukawa
terms
\begin{align}\label{eqn:Yukbif}
W &= 2 m_1 \tilde{Q}^{bc}  
Q_{b c}  - 2 \e^{c c'}  \tilde{Q}_{bc} \Phi_2^{bb'} Q_{b'c'} - 2
\e^{b b'}  \tilde{Q}_{bc} \Phi_3^{cc'} Q_{b' c'} 
\end{align}
of the bifundamental hyper of mass $m_1$. Here we made use of the identity
$\sigma_G^{-1} T \sigma_G = -T^t$. 

Remark that the
identifications in equation~(\ref{eqn:trifbifid}) reduce the $SU(2)$
R-symmetry for the trifund to that of the bifund, while identifying
the gauge symmetry of the first $SU(2)$ gauge group with the enhanced
flavor symmetry of 
the bifund. Also notice that the mass-term breaks the
enhanced flavor symmetry of the bifund.

Let us continue by Higgsing the second gauge group  by setting 
$(\Phi_2)^{b}_{~b'} = m_2 (\sigma_3)^{b}_{~b'}$. The single
bifundamental hyper turns into two fundamental hypers
\be\label{eqn:biffundid}
 Q_{(k)c}  &\equiv& Q_{kc}    \\
\tilde{Q}_{(k)c} & \equiv& \tilde{Q}_{kc},  \notag 
\ee
where a subscript $(.)$ refer to a flavor index. The Yukawa terms can be repackaged as
\begin{align}
W
=&~  m_1  \, (\sigma_3)^{g}_{~f} \, \delta^{l}_{~k} \, Q^{(f)(k)c}
Q_{(g)(l)c}+ m_2 \, \delta^{g}_{~f} \,  (\sigma_3)^l_{~k} \, 
Q^{(f)(k)c} Q_{(g)(l)c} \\
&+ \e^{fg} \, \e^{kl}
Q_{(f)(k)c} \Phi_3^{cc'} Q_{(g)(l)c'} \notag,
\end{align}
if we furthermore make the identifications $Q = Q_{(f=1)}$ and
$\tilde{Q} = Q_{(f=2)}$. This superpotential describes two fundamental
hypers whose flavor symmetry enhances to $SO(4)$ when the masses are
turned off.

As a  consistency check let us Higgs both bifundamental gauge groups by
setting  
$(\Phi_2)^{b}_{~b'} = m_2 (\sigma_3)^{b}_{~b'}$ and
$(\Phi_3)^{c}_{~c'} = m_3 (\sigma_3)^{c}_{~c'}$ in
equation~(\ref{eqn:Yukbif}). This results in the Yukawa terms   
\begin{align}
W=&~ m_1  \, \delta^l_{~k} \, \delta^n_{~m} \,
(\sigma_3)^g_{~f} \, Q^{(f)(k)(m)} 
Q_{(g)(l)(n)}+ m_2 \, (\sigma_3)^{l}_{~k} \, \delta^n_{~m} \, \delta^g_{~f} \,
Q^{(f)(k)(m)} Q_{(g)(l)(n)} \\
& + m_3 \,  \delta^l_{~k} \, (\sigma_3)^n_{~m} \, \delta^g_{~f} \,
Q^{(f)(k)(m)} Q_{(g)(l)(n)},\notag 
\end{align}
corresponding to the superpotential of eight half-hypers with
a diagonal mass matrix with eigenvalues $\pm m_1 \pm m_2 \pm
m_3$, as expected.

\section{Contour integrands for Sicilian quivers}\label{app:contourintegrands}

In this appendix we summarize the contour integrand formulae for the
Sicilian quiver gauge theories that we encountered in the main text. More
details on instanton counting can be found in \cite{HKS}. 

\subsection*{$Sp(1)-SO(4)$ bifundamental full hypermultiplet}

The contour integrand for the massive full $Sp(1)-SO(4)$ bifundamental
hyper is  
\be \label{zSpSOdoublebifund}
 \mathbf{z}_{\rm{2bif}, k_1, k_2}^{Sp(1)-SO(4)}(\phi, \psi, \m, a, b_1,b_2)
 &=& \left( \prod_{l=1}^{2} \Delta_1 ( \m \pm  b_l) \right) \Delta_2 ( \m  \pm a)  P_2 (\m)^{\chi_\phi} \\
 && \times \left( \frac{\Delta(\m - \e_- ) \Delta(\m + \e_-) }{\Delta(\m + \e_+) \Delta(\m - \e_+)} \right)  \left( \frac{\Delta_2 (\m - \e_-)\Delta_2 (\m + \e_-)}{\Delta_2 (\m + \e_+)\Delta_2 (\m - \e_+)} \right)^{\chi_\phi} , \nn
\ee
where the $Sp(1)$ instanton parameter $k_1 = 2n_1 + \chi_\phi$, the
deformation parameters $\e_\pm = \frac{\e_1 \pm \e_2}{2}$ and
$\pm$ is an abbreviation for a product over both terms. Furthermore,
$\m$ is the physical mass parameter and
\begin{align*}
 \Delta_1 (x) &= \prod_{i=1}^{n_1} (\phi_i^2 - x^2) \\
 \Delta_2 (x) &= \prod_{j=1}^{k_2} (\psi_j^2 - x^2) \\
 \Delta (x) &= \prod_{i, j=1}^{n_1, k_2} \left( (\phi_i + \psi_j )^2 - x^2 \right) \left( (\phi_i - \psi_j )^2 - x^2 \right) \\
 P_2 (x, b) &= \prod_{l=1}^{n_2 } (b_l^2 - x^2). 
\end{align*}

\subsection*{$Sp(1)-SO(4)$ single bifundamental half-hypermultiplet}

The $Sp(1)-SO(4)$ double bifundamental
integrand~(\ref{zSpSOdoublebifund}) becomes a complete square 
when the bifundamental mass vanishes. The integrand
for the bifundamental half-hyper is thus simply 
\be
\mathbf{z}_{\rm{bif}, k_1, k_2}^{Sp(1)-SO(4)}(\phi, \psi, a, b_1,b_2) &=& \,
 \prod_{i=1}^{n_1} \left( \phi_i^2 - b_1^2 \right)\left( \phi_i^2 -
   b_2^2 \right) \prod_{j=1}^{k_2} \left( a^2 - \psi_j^2 \right)
 \frac{\Delta(\e_-)}{\Delta(\e_+ )}  \left( b_1 b_2   \frac{\Delta_2
     (\e_-)}{\Delta_2 (\e_+)}\right)^{\chi_\phi}\,. \nn
\ee
The factor $\Delta(\e_+)$ in the denominator cannot be canceled by a
contribution from the gauge multiplets, and therefore brings in additional
poles.  

\subsection*{$U(2)^3$ trifundamental full hypermultiplet}

Starting from the equivariant index
\be\label{eq:equivindextrifu2}
 \Ind_{\mathbf{T}} = \int_{\IC^2} \Ch_{\mathbf{T}} (\CE_{U(N_1)}
 \otimes \CE_{U(N_2)} \otimes \CE_{U(N_3)} \otimes \CL \otimes M ) \,
 \Td_{\mathbf{T}} (\IC^2)\, ,
\ee
we obtain the contour integrand
\be
\mathbf{z}_{k_1, k_2, k_3}^{(N_1, N_2, N_3)} &=& 
\prod_{i, m, n}^{k_1, N_2, N_3}(\phi_{1, i} + b_m + c_n + \m)
\prod_{j, l, n}^{k_2, N_1, N_3} (\phi_{2, j} + a_l + c_n + \m)  
\prod_{k, l, m}^{k_3, N_1, N_2} (\phi_{3, k} + a_l + b_m + \m)  \nn \\
& \times & \prod_{i, j, n}^{k_1, k_2, N_3}\frac{(\phi_{1, i} + \phi_{2, j} + c_n - \e_- + \m)(\phi_{1, i} + \phi_{2, j} + c_n + \e_- + \m)}
{(\phi_{1, i} + \phi_{2, j} + c_n + \m +\e_+)(\phi_{1, i} + \phi_{2, j} + c_n + \m - \e_+)} \nn \\
& \times & \prod_{i, k,m}^{k_2, k_3, N_1}\frac{(\phi_{1,i} + \phi_{3, k} + b_m - \e_- + \m)(\phi_{1, i} + \phi_{3, k} + b_m + \e_- + \m)}
{(\phi_{1, i} + \phi_{3, k} + b_m + \m + \e_+)(\phi_{1, i} + \phi_{3, k} + b_m + \m - \e_+)} \\
& \times & \prod_{j, k, l}^{k_2, k_3, N_1}\frac{(\phi_{2, j} + \phi_{3, k} + a_l - \e_- + \m) (\phi_{2, j} + \phi_{3, k} + a_l + \e_- + \m)}
{(\phi_{2, j} + \phi_{3, k} + a_l +\m + \e_+)(\phi_{2, j} + \phi_{3, k} + a_l + \m - \e_+)}  \nn \\
& \times & \prod_{i, j, k}^{k_1, k_2, k_3} \frac{(\phi^{123}_{ijk} + \e + \m)(\phi^{123}_{ijk} + \e_1 - \e_2 + \m)(\phi^{123}_{ijk} - \e_1 + \e_2 + \m)(\phi^{123}_{ijk} + \m)^4 (\phi^{123}_{ijk} + \m - \e)}
{(\phi^{123}_{ijk} + \e_1 + \m)^2 (\phi^{123}_{ijk} + \e_2 + \m)^2 (\phi^{123}_{ijk} - \e_1 + \m)^2 (\phi^{123}_{ijk} - \e_2 + \m)^2} \nn
\ee
where $\phi^{123}_{ijk} = \phi_{1, i} + \phi_{2, j} + \phi_{3, k}$ and $\e = \e_1 + \e_2$ and $\e_\pm = \frac{\e_1 \pm \e_2}{2}$. 
The contour integrand for the massive full $SU(2)$ trifundamental
hypermultiplet is found by setting $N_{1/2/3} = 2$. If we set the
instanton parameter $k_3 = 0$, we recover the contour integrand for
two copies of the bifundamental of mass $\pm c$. 

The one-loop contribution to the Nekrasov partition function 
\be 
 Z_{\textrm{1-loop}}^{\rm 2trif} &=& \prod_{i, j = 1}^\infty \prod_{l,
   m, n=1}^2 \left(a_l + b_m +  c_n + \mu + \e_+ + i \e_1 + j \e_2 \right)^{-1} \nn 
\ee
is obtained from the perturbative contribution to the equivariant
index (\ref{eq:equivindextrifu2}).

\end{document}